\documentclass[a4paper,11pt]{article}
\pdfoutput=1 
\usepackage{jheppub2} 
\usepackage{bbm,bm,graphicx,mathtools,color,hyperref,slashed}
\usepackage{amssymb,amsmath}
\usepackage{comment}
\usepackage[all]{xy}
\usepackage[dvipsnames]{xcolor}

\graphicspath{{./Figures/}}

\newcommand{\tr}{\mathrm{tr}}

\newcommand*{\half}{\textstyle{ \frac{1}{2}}}

\newcommand{\del}{\partial}
\newcommand{\e}{\mathcal{E}}

\newcommand{\bra}{\langle}
\newcommand{\ket}{\rangle}
\newcommand{\cg}{\textnormal{\textsl{g}}}

\preprint{.}

\title{Quantum Hamilton-Jacobi Theory, Spectral  Path Integrals and Exact-WKB
analysis}

\author{Mustafa T\"{u}re,}
\author{Mithat \"{U}nsal}

\affiliation{Department of Physics, North Carolina State University, Raleigh, NC 27607, USA}

\emailAdd{mture@ncsu.edu, unsal.mithat@gmail.com}

\abstract{
We propose a new way to perform path integrals in quantum mechanics   by using a quantum version of Hamilton-Jacobi theory. 
In classical mechanics, Hamilton-Jacobi theory is a powerful formalism,   however, its utility is not explored in quantum theory beyond  approximation schemes.  The canonical transformation 
 enables one to set the new Hamiltonian to constant or zero, but  keeps  the information about solution in Hamilton's characteristic function. To benefit from this    
 in quantum theory, one must work with a formulation in which classical Hamiltonian  is used.  This uniquely  points to phase space path integral. However, the main variable in HJ-formalism is energy, not time. Thus, we are led to consider  Fourier transform of  path integral, spectral path integral, $\tilde Z(E)$. 
The evaluation of path integral reduces to determining the quantum Hamilton's characteristic functions (which can be achieved via an asymptotic analysis), and a discrete sum  over the quantum period lattice, generalizing  Gutzwiller's sum.
 
}

\begin{document}
\maketitle
%--------------------------------------------------------------------------------------

\begin{quote} 
{\it "I like to find new things in old things." } \quad Michael Berry
\end{quote}

\section{Introduction}

Classical mechanics is to quantum mechanics what 
 geometric optics is to wave optics.  In the cleanest form of the correspondence principle, Hamilton-Jacobi formulation of classical mechanics plays a prominent role. Let us briefly remind its well-known version.  For a time-independent Hamiltonian, using $\Psi (q, t) \sim e^{i(-Et + W(q))/\hbar} $, Schr\"odinger equation becomes the non-linear Riccati equation, 
   \begin{align} 
 \frac{1}{2} \left(\frac{d W}{d q}\right)^2 + V(q) -  \frac{i \hbar  }{2}  \frac{d^2 W}{d q^2} = E
 \label{Riccati1}
 \end{align}
For $\hbar \neq 0$, \eqref{Riccati1} is an exact representation of the Schr\"odinger equation. For  $\hbar = 0$,   \eqref{Riccati1}  is an exact representation of classical mechanics, it is  the equation for Hamilton's characteristic function $W(q)$, which carries the same information as Newtonian, Lagrangian, or Hamiltonian formulation. 
 At  infinitesimal  $\hbar$,  $W(q, \hbar)$ in  \eqref{Riccati1} should be viewed as the quantum generalization of the Hamilton characteristic function.

Riccati equation is the starting point of the exact WKB formalism \cite{ 
 dillinger_resurgence_1993,DDP2, balian_discrepancies_1978,Voros1983, Silverstone, 
 aoki1995algebraic, AKT1, sueishi_exact-wkb_2020}.   It can be converted to a recursive equation the solution of which is given in terms of asymptotic series in $\hbar$.   Exact-WKB is the study of a differential equation in complexified  coordinate space    ($q \in \mathbb C$)  by using  
resurgence theory \cite{Ec1,sauzin2014introduction}, and Stokes graphs.      
  Classical data about the potential dictates the Stokes graph, and  demanding monodromy free condition from  the WKB-wave function and demanding that it vanishes at infinities for bounded potentials lead to the exact quantization conditions.\footnote{There are two conditions here. Monodromy free condition   leads to perturbative quantization and discreteness of energy. Vanishing at the wedges at infinity leads to the full non-perturbative quantization condition. }

Of course, this is a very elegant formalism but it is also clear that this line of reasoning does not take advantage of the  full Hamilton-Jacobi theory.  
In classical mechanics,   the  power of  Hamilton-Jacobi theory stems from the ability to select a canonical transformation to new coordinates which are either constants or cyclic. In particular, one can even choose a  generating function such that the new Hamiltonian is zero,  or a constant $c$, 
\begin{align}
\tilde H=0 \quad {\rm or}  \quad \tilde H=c
\end{align} 
 The whole solution of the classical system is in the reduced action, $W(q, E)$.   What is the implication/benefit of classical canonical transformations in the context of quantum mechanics? 
% More specifically,  
% \begin{quote}
% What does it mean to solve a quantum theory by using a 
%  canonical transformations, where the classical Hamiltonian can  even be set to  zero? 
% \end{quote}

  To explore the answer to this question in {\it quantum theory},  
 we must work in a formulation  of quantum mechanics in which {\it classical  Hamiltonian} enters the story.  
 This uniquely points  us to  work with the phase space path integral. 
 One may be tempted to think that one should work with the phase space    path integral  in the standard form:
 \begin{equation}
Z(T)= \tr e^{-\frac{i}{\hbar} \widehat HT} = \int \mathcal{D}q\mathcal{D}p \; e^{\frac{i}{\hbar} \int_0^T (p \dot q -  H) dt  }  
\label{partition}
\end{equation} 
The trace implies that we need to integrate over paths satisfying periodic boundary conditions $x(T) = x(0),  \; p(T)= p(0)$, returning to themselves at fixed time $T$.   Such paths can be rather wild, unearthly, and the energy $E$ can take any value, even infinity so long as it is 
periodic.\footnote{The same is also true 
in  configuration space path integral  with uses classical Lagrangian,   
$ \int \mathcal{D}q \; e^{\frac{i}{\hbar} \int_0^T {\cal L} dt } $. It also does not matter if we are considering 
Minkowski time or Euclidean time, where the latter corresponds to thermal partition function $Z(\beta)$}.
However, the key player in classical Hamilton-Jacobi theory,  Hamilton's characteristic function $W(E, q)$ is a function of $E$. 
It would be more natural to work with all paths not at a fixed time $T$, but at a fixed energy   $E$.   

In some way, we would like to perform phase space  path integrals not at a fixed $T$ 
(where $E$ can take arbitrarily large values),  but  at fixed $E$ (where $T$ can take arbitrarily large values). 
Therefore, it is more natural to work with the Fourier  transform of the path integral,  which is a function of $E$. 
     \begin{equation} 
\widetilde Z(E) = \frac{i}{\hbar}  \int dT \; e^{\frac{i}{\hbar} E T}  Z(T) = \tr  \left(\frac{1}{E- \widehat H}\right) \equiv G(E)
\label{spectral0}
\end{equation}  
This is nothing but the resolvent for the original Hamiltonian operator $\widehat H $, and it  is simply  related to spectral determinant  
\begin{equation}
    D(E) := \det( \widehat H-E),   \quad   -\frac{\del}{\del E}\log D(E) = G(E)
    \label{spec-det}
\end{equation} 
 In our context, it is more natural to view both of these as phase space path integrals which are functions of $E$,  i.e.,   {\it spectral path integrals},  and we will use this terminology interchangeably with  resolvent and spectral determinant.  

 It is  the spectral path integral $\widetilde Z(E)$ 
\eqref{spectral0}, not   the original $Z(T)$   \eqref{partition}, that allows us to explore the implication of the  Hamilton-Jacobi formalism for the path integrals in phase space.  Physically, the main advantage of   $\widetilde Z(E)$, as we will demonstrate,   is that the set of paths that contribute path integral is discrete infinity, and countable. This is  unlike the one that enters in configurations space path integral $Z(T)$, which is continuous infinity.

In essence, we  generalize  the classical canonical transformations of the Hamilton-Jacobi to a quantum canonical transformation in the context of phase space path integral. The classical reduced actions are promoted  to  quantum ones:
\begin{align}
W_{\gamma}(E)  \rightarrow  W_{\gamma}(E, \hbar)
\end{align}
In classical mechanics, only the classically allowed cycles $\gamma$  contribute to the equations of motions and dynamics. 
 What is the set of cycles that enter the quantum theory? \footnote{This question, in configuration space path integral formulation,  is equivalent to 
which saddles contribute to the path integral?  What is the role of generic complex saddles?  }  

  It turns out this question has a sharp answer. It is given in terms of what we will refer to as  {\it vanishing cycles}.  These are the cycles for which as $E$ is varied, {\it two or more} turning points coalesce at some critical $E$.   The critical $E$ are associated with 
  separatrices in classical mechanics of the potential problems for $V(q)$ and $-V(q)$. 
   The classically allowed vanishing cycles will be called perturbative (or classical) cycles  $\gamma_{i}$, and 
  classically forbidden vanishing cycles will be called non-perturbative (or dual)  cycles $\gamma_{d,j}$.  
   \begin{align}
   \Gamma = \{ \gamma_{i},  \;  \gamma_{d, j}, \quad   i=1, \ldots, N,  \quad   j=1, \ldots, M  \}
   \label{cycles}
\end{align}

The spectral path integrals will be expressed in terms of  quantum generalizations of Hamilton's characteristic functions on these cycles,  $e^{\frac{i}{\hbar}W_{\gamma_{i}}(E, \hbar)} \equiv A_i$ and  $  e^{\frac{i}{\hbar}W_{\gamma_{d, i}}(E, \hbar)}  \equiv B_i$, also called Voros multipliers. 
  There are important advantages gained from the spectral path integral. 
\begin{itemize}
\item Upon Hamilton-Jacobi canonical transformation, the spectral phase space path integral  $\widetilde Z(E)$  becomes a discrete sum involving the quantum version of the reduced actions $W_{\gamma}(E, \hbar)$ associated with independent vanishing cycles,  $\gamma \in \Gamma$.  

\item The discrete sums  in terms of vanishing cycles can be performed analytically.  The remarkable fact is that 
the path integral sum produces the exact quantization conditions that is obtained in the exact-WKB analysis, 
by the study of differential equations in complex domains. 
Our construction, based on quantum Hamilton-Jacobi, provides a streamlined derivation of the proposal in \cite{sueishi_exact-wkb_2020,sueishi_exact-wkb_2021,kamata_exact_2023}. 

\item The spectral determinant and  spectral partition function  can be written as 
\begin{align}
D(E)= D_{\rm p}(E) D_{\rm np}(E), \qquad 
\tilde Z(E)=  \tilde Z_{\rm p}(E) +   \tilde Z_{\rm np}(E),  
\end{align}
where  subscripts $ ({\rm p}, {\rm np}) $  denote  the sums over perturbative and non-perturbative vanishing cycles, respectively. 
The zeros of $D(E)$ gives the spectrum of quantum theory. 
This shows that the quantum spectrum of the theory can be explained in terms of properties of the classical cycles, with both  p and np cycles included, answering Gutzwiller's fundamental question \cite{gutzwiller_periodic_1971}. (See also \cite{Witten:2010zr}).

\end{itemize}

\begin{figure}[t]
    \centering
    \includegraphics[scale=0.15]{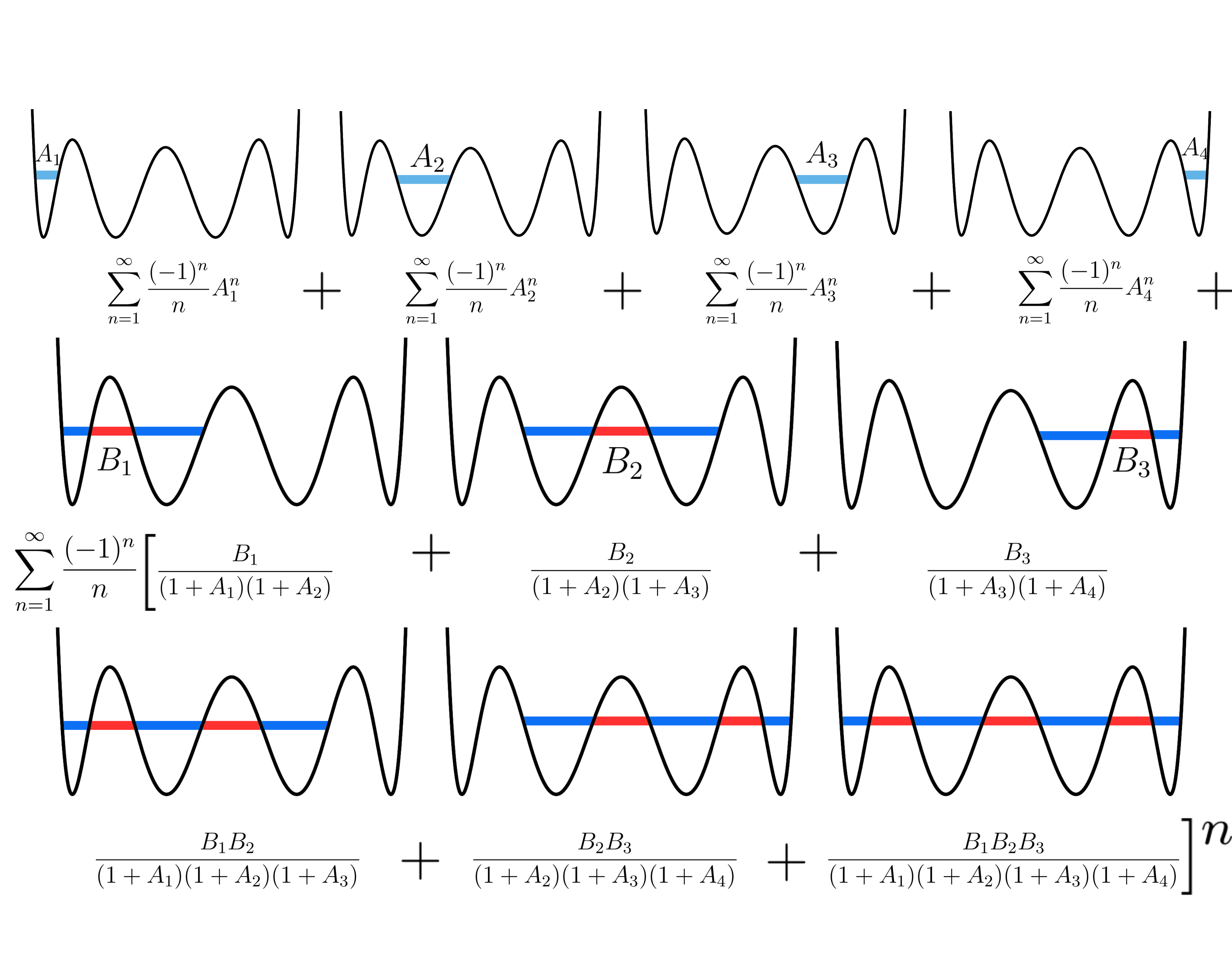}
    \vspace{-1cm}
    \caption{Sum over all periodic orbits at fixed energy $E$. The first line is the sum of all perturbative periodic orbits, where $ A_i(E, \hbar) = e^{i W_{\gamma_i}(E, \hbar)}$.  The rest is the sum over all non-perturbative periodic orbits. NP periodic orbits are dressed 
    with infinite copies of perturbative periodic orbits. NP orbits also serve as connectors between perturbative periodic orbits.  
    }
    \label{fig:orbits}
\end{figure}

 As an example, the spectral partition function for a 4-well potential problem reduces to the discrete sum 
over the orbits shown in Fig.~\ref{fig:orbits}. The first line is the sum over perturbative vanishing cycles. The sum gives 
$ D_{\rm p}(E) $.  The second and third are the sum over the 
  nonperturbative cycles and they add up to $ D_{\rm np}(E) $.  
  $ D(E)=0 $ gives the quantum spectrum of the theory.  

\vspace{3mm}
\noindent
{\bf Relation to other works}    

The central theme of the exact WKB analysis is the study of differential equations in complex domains, by using asymptotic analysis, resurgence and Stokes graphs.  
The philosophy of this work is complementary. It is the study of the spectral 
phase space path integral by using the (quantum) Hamilton-Jacobi formalism.  
Although methods are  different, both yield exact quantization conditions. 
There is ultimately an overlap 
of the two formalisms, because exact quantization conditions are expressed in terms of 
Hamilton's characteristic functions (reduced actions or Voros symbols).  
However, we believe the path integration demystifies the origins of the quantization conditions, as sum over p/np  periodic orbits, while in the exact WKB, the same condition arises from the normalizability of the analytically continued WKB wave function $\Psi_{\rm WKB}(q)$  as $q \rightarrow \infty$.  The remarkable fact is that both of these expressions are given in terms of  $e^{\frac{i}{\hbar}W_{\gamma} (E, \hbar)} $ for the p/np cycles in the problem.

Below, we mention some important developments on the exact WKB and quantization conditions. 
Exact  quantization condition capturing all multi-instanton effects  was  conjectured by \cite{Zinn-Justin:2004vcw, Zinn-Justin:2004qzw,  jentschura_multi-instantons_2010,jentschura_multi-instantons_2011} as generalized Bohr-Sommerfeld quantization, which is based on classically allowed
(P) and classically forbidden (NP) cycles.  The conjecture was proven  \cite{dillinger_resurgence_1993,DDP2} following the work of \cite{balian_discrepancies_1978,Voros1983, Silverstone}, and using resurgence theory \cite{Ec1} in some special cases.

Another important result is derived for genus-1 potential problems.  The  full non-perturbative expression for energy eigenvalues, containing all orders of perturbative and  non-perturbative  terms, may be generated directly from the perturbative expansion about the perturbative vacuum \cite{Dunne:2013ada, Dunne:2014bca, Alvarez3, Cavusoglu:2023bai}.   This fact is quite remarkable and its generalization to higher genus potentials is an open problem.

  The WKB connection formulae, together with the condition of monodromy-free wavefunction and its vanishing at real infinities lead one to find the spectral determinant of a quantum mechanical system, hence the exact quantization condition for general 
 $ N$-well systems. 
  Consequently, an explicit connection between the exact WKB theory and the path integral was made in 
 \cite{sueishi_exact-wkb_2020,sueishi_exact-wkb_2021,kamata_exact_2023}, also the relation to 
 the pioneering work of  Gutzwiller \cite{
 gutzwiller_periodic_1971} is pointed out.   Our work differs from  \cite{sueishi_exact-wkb_2020,sueishi_exact-wkb_2021,kamata_exact_2023} in that it  provides a derivation of the correspondence starting solely from the path integral, and the crucial use of 
 classical/quantum Hamilton-Jacobi transformation. This perspective provides new insights into the nature of path integral and did not appear in earlier studies.

 The  sum over vanishing cycles $\Gamma$  gives a generalization of the Gutzwiller summation formula. The  
Gutzwiller summation formula in its original form  
is  phrased in terms of cycles that enter classical mechanics (prime periodic orbits) \cite{gutzwiller_periodic_1971, MVBerry_1977}.   In quantum mechanical path integral for general potential problems,  vanishing cycles also include non-perturbative (tunneling or instanton) cycles. 
  For   recent developments in semi-classics and exact WKB in quantum mechanical systems, see \cite{ Nekrasov:2018pqq, Song:2024ozi, Basar:2013eka,Gahramanov:2015yxk,Behtash:2015kna,Dunne:2016jsr,Dunne:2016qix,Kozcaz:2016wvy,Basar:2017hpr,Behtash:2015loa,Codesido_2018,Behtash:2018voa,Sulejmanpasic:2016fwr,Dunne:2020gtk,Iwaki1, Kamata:2023opn, Kamata:2024tyb, Bucciotti:2023trp}.  For the relation between  ${\cal N}=2$ supersymmetric gauge theory and exact WKB, see\cite{Gaiotto:2009hg, Kashani-Poor:2015pca, Basar:2015xna, Ashok:2016yxz, Yan:2020kkb, Hollands:2019wbr, Grassi:2021wpw, Grassi:2022zuk}

\section{Spectral Determinant as Spectral (Phase Space) Path Integral} 
We first remind briefly of the relation between the resolvent,  spectral determinant, and phase space path integral.  These are quantities  that possess complete information about the energy spectrum of the quantum system. 
As in \cite{muratore-ginanneschi_path_2003, Schulman:1981vu},
we start with the usual definition of the propagator
\begin{equation}
    K\left(q,t\mid q',t'\right) = \langle q|e^{-\frac{i}{\hbar}H(t-t')}|q'\ket, \qquad t\geq t',
    \label{prop}
\end{equation}
and $K=0$ for $t<t'$. The propagator \eqref{prop}  works as the kernel of the Schr\"odinger equation,
\begin{equation}
    \left(i\hbar \frac{\del}{\del t} - H\right)K\left(q,t\mid q',t'\right) = -i\delta(q-q')\delta(t-t').
\end{equation}
Since we are only considering forward propagation in time, we can rewrite $K$  as
\begin{equation}
    K\left(q,t\mid q',t'\right) := G(q,q'|T)\Theta(t-t')
\end{equation}
where $T = t-t'$ and $ \Theta(t-t')$ is step function.
Assuming that the corresponding Hilbert space is spanned by a complete set of energy eigenstates $\{|\alpha\ket\}$ of the Hamiltonian operator $H$, we can express the propagator as
\begin{equation}
    K(q,T\mid q',0) = i\sum_{\alpha}\bra q|\alpha\ket\bra\alpha|q'\ket e^{-\frac{i}{\hbar}E_{\alpha}T}
\end{equation}
The propagator has information on the spectrum $E_{\alpha}$,  and is a function of time $T$.

It is more convenient  to define a spectral  function as a function of  energy via the Fourier transform  
\begin{align}
   \label{Fourier}
    iG(q,q'\mid E) &= \int_{-\infty}^{\infty}\frac{dT}{\hbar}e^{\frac{i}{\hbar}ET}G(q,q'\mid T)\Theta(T) \cr
    &=\int_0^{\infty}\frac{dT}{\hbar}e^{\frac{i}{\hbar}ET}G(q,q'\mid T).
\end{align}
In particular, when we use path integrals,  the Fourier transform  will allow us to work with paths with fixed $E$, rather than paths with fixed $T$. This step will also be crucial in carrying over Hamilton-Jacobi to quantum theory. 
Using  \eqref{Fourier},  we have
\begin{align}
    G(q,q'\mid E) &= \sum_{\alpha}\overline{\psi}_{\alpha}(q')\psi_{\alpha}(q)\int_0^{\infty}\frac{dT}{ i \hbar}e^{\frac{i}{\hbar}(E-E_{\alpha})T}   \cr 
    &= \sum_{\alpha}\overline{\psi}_{\alpha}(q')\psi_{\alpha}(q)\frac{1}{E-E_{\alpha}} = \bra q\mid\frac{1}{E-H}\mid q'\ket. 
\end{align}
To obtain the second line, we performed the  integration  over $T$  by adding  $E$ a small  imaginary complex term 
$E \rightarrow E+  i  \epsilon$  for convergence,  and then  let $ \epsilon \rightarrow 0 $ in the final result. This is the resolvent associated with the Hamiltonian operator, and satisfies:

\begin{equation}
    (H-E)G(q,q'\mid E) = -\delta(q-q'),
\end{equation}
Therefore,  one maps the energy spectrum of the system to the poles of the trace of the resolvent
\begin{equation}
    G(E):= -\int dq_0 G(q_0,q_0\mid E) = \text{Tr}\frac{1}{H-E}.\label{trace}
\end{equation}

A related  important object that encodes spectral data is the Fredholm determinant
\begin{equation}
    D(E) := \det(H-E)
\end{equation}
such that
$D(E) = 0$ is the quantization condition for a system with Hamiltonian $H$. Observe that the resolvent and the Fredholm determinant are related to each other as
\begin{equation}
    -\frac{\del}{\del E}\log D(E) = G(E).
    \label{res-det}
\end{equation}
Using the inverse Fourier/Laplace transform of $G(E)$ one gets
\begin{equation}
 G(E)=   -\frac{\del}{\del E}\log D(E) = \frac{i}{\hbar}\int_0^{\infty}dT e^{\frac{i}{\hbar}ET}G(T)
\end{equation}
since $G(T)$ is nothing but the trace of the propagator
\begin{equation}
    G(T) = \int dq_0 \bra q_0\mid e^{-\frac{i}{\hbar}HT}\mid q_0\ket = \int dq_0\int_{q(0)=q(T)=q_0}\mathcal{D}[q(t)]\mathcal{D}[p(t)]e^{\frac{i}{\hbar}\int_0^Tdt\left(p\Dot{q}-H\right)}.
\end{equation}

Putting everything back into \eqref{res-det}, one gets the relationship
\begin{equation}
  G(E)=  -\frac{\del}{\del E}\log D(E) = \frac{i}{\hbar}\int_0^{\infty}dT\int dq_0\int_{q(0)=q(T)=q_0}\mathcal{D}[q(t)]\mathcal{D}[p(t)]e^{\frac{i}{\hbar}\left(ET-\int_0^THdt + \oint pdq\right)}.\label{resolvent}
\end{equation}
At first sight, this equation is easy to derive and looks pretty simple. It is the relation between the spectral resolvent and phase space in path integral. 
In what follows, we present a formulation that represents path integral in terms of cycles in phase space.

\section{Classical and Quantum Canonical Transformations}
 We first discuss the  Hamilton-Jacobi formalism  in classical mechanics, and next, we discuss its  implementation to quantum theory. 
For any classical system, one can consider a canonical transformation $(q,p)\to(Q,P)$ defined by the type-2 generating function $G_2(q,P)$, such that
\begin{equation}
    Q \equiv \frac{\del G_2(q,P)}{\del P}, \qquad p \equiv \frac{\del G_2(q,P)}{\del q}.
\label{ct}
\end{equation}
One can choose the action to be the generating function up to a constant
\begin{equation}
    G_2(q,P)= S(q,P) + C=
    \int (p\Dot{q}-H)dt + C
\end{equation}
then we see that the new Hamiltonian $\Tilde{H}$ in terms of the new coordinates is identically zero,
\begin{equation}
    \Tilde{H}(Q,P,t) = H + \frac{\del S}{\del t} = 0
    \label{ct_hamil}
\end{equation}
Therefore, the new coordinates are constants of motion, i.e., 
 their equations of motion are trivial:
\begin{equation}
    \frac{\del \Tilde{H}}{\del Q} = -\Dot{P} = 0, \qquad \frac{\del \Tilde{H}}{\del P} = \Dot{Q} = 0.
\end{equation}
Writing everything in terms of the old coordinates,  \eqref{ct_hamil} gives the Hamilton-Jacobi equation
\begin{equation}
    \frac{\del S}{\del t} = -H\left(q,\frac{\del S}{\del q},t\right)
    \label{HJ}
\end{equation}
If one considers a system with a time-independent Hamiltonian, we can separate the variables of $S$ as
\begin{equation}
    S = -Et + W(q) = -Et + \int^qp(q',E)dq' 
    \label{SOV} 
\end{equation}
Here, the time-independent term,  $W(q)$  is called  Hamilton's characteristic
function or reduced action. 
By substituting \eqref{SOV} into 
 Hamilton-Jacobi equation  \eqref{HJ}, we obtain the equation for  $W(q)$.  
\begin{equation}
    H\left(q,\frac{\del W}{\del q}
    \right) = E
    \label{energy}
\end{equation}
which defines the classical trajectories as the level sets of the Hamiltonian. One can choose the new coordinates and momenta as the initial time $Q\equiv-t_0$ and the energy $P\equiv E$, both being the constants of motion. Then, \eqref{energy} implies the form of the old momentum $p$ trajectories to be
\begin{equation}
    p^2(q,E) = 2\left(E-V(q)\right).
\end{equation}

\noindent
\subsection{Quantum Hamilton-Jacobi Transformation}
To carry out a similar implementation to quantum mechanics, we need to promote the action to be the "quantum" action. Assume that there exists a quantum generating function $S_Q(q,\mathbf{P};\hbar)$ that results in the canonical transformation
\begin{equation}\label{QCT}
    (q,p_Q) \to (\tau_Q,\Tilde{E}) \equiv (\mathbf{Q},\mathbf{P}).
\end{equation}
For systems with time-independent Hamiltonian, we may again take the generating function to be the quantum action with separated variables
\begin{equation}
    S_Q = -\Tilde{E}t + W_Q(q,\Tilde{E},\hbar) = -\Tilde{E}t + \int^qp_Q(q',\Tilde{E},\hbar)dq'.
\end{equation}
However, this time the trajectories satisfy the quantum version of the  Hamilton-Jacobi equation which is a partial differential (Riccati) equation
\begin{equation}
    -\frac{\del S_Q}{\del t} = \frac{1}{2}\left(\frac{\del S_Q}{\del q}\right)^2 + V(q) - \frac{i\hbar}{2}\frac{\del^2 S_Q}{\del q^2}
    \label{Riccati}
\end{equation}
equivalent to the Schr\"odinger equation. This implicitly defines the quantum momentum function $p_Q$ as a function of $q$ and $\Tilde{E}$ satisfying
\begin{equation}
    2\left( \Tilde{E}-V(q)\right) + i\hbar\frac{\del p_Q(q,\Tilde{E})}{\del q} = p_Q^2(q,\Tilde{E})
\end{equation}
Everything defined above is equivalent to classical mechanics in the limit $\hbar\to0$ so one can see the above equation as a defining equation for the quantum perturbations to the classical momentum.

One can find a formal asymptotic series solution to the Quantum Hamilton-Jacobi equation for the quantum Hamilton's characteristic function $W_Q$
\begin{equation}
    W_Q(q,E) = \sum_{n=0}^{\infty}\int dq'\frac{\rho_n}{p^{3n-1}}\hbar^n
\end{equation}
where $\rho_n(q,E)$ satisfies the recursion relation \cite{Fischbach_2019},
\begin{equation}
    \rho_{n+1} = \frac{i}{2}\left(2(E-V)\frac{\del\rho_n}{\del q}+(3n-1)V'\rho_n\right)-\frac{1}{2}\sum_{k=1}^n\rho_k\rho_{n+1-k}
\end{equation}
with $\rho_0=1$, $\rho_1=-\frac{i}{2}V'$ and $p = \sqrt{2(E-V)}$ is the classical momentum. When periodic boundary conditions are imposed $q(T)=q(0)=q_0$, we get the quantum reduced action as an asymptotic series in $\hbar$,
\begin{align}
    W_{Q,\Gamma}(E) &= \sum_{n=0}^{\infty}\oint_{\Gamma}dq\frac{\rho_n}{p^{3n-1}}\hbar^n  \cr
   &=  \sqrt{2} \left(  \oint_{\Gamma} \sqrt{(E-V)} dq  - \frac{ \hbar^2 }{2^6}  \oint_{\Gamma}  \frac{ (V')^2 } { (E-V)^{5/2} } dq   + O(\hbar^4) \right)
   \label{Red-ac}
\end{align}
The integral at each order $n$ takes nonzero values only on nontrivial 1-cycles of the family of compact Riemann surfaces defined by the (hyper)elliptic curve
\begin{equation}
    \e: p^2 = 2(E-V(q)).
\end{equation}
For a given holomorphic $V(q)$ and $E\in\mathbb{R}$, the complex curve $\e_E$ is isomorphic to a compact Riemann surface X of genus-g whose singular points are when $E=V_c$, (critical values of V), and branch points for the 2 complex sheets are $p=0\implies E=V(q)$. Then the set of differentials $\left\{\frac{\rho_n}{p^{3n-1}}dq\right\}^{\infty}_{n=0}$ are elements of the first cohomology group $H^1(X,\mathbb{C})$ of dimension 2g. One can choose the basis of the first homology group $H_1(X,\mathbb{C})$ to be the cycles $\{a_i,b_i\}_{i=1}^{\cg}$ that encircle the turning points, i.e. the branch points. It is convenient to choose $a_i$ to be the cycles that correspond to the trajectories of classically allowed regions and choose $b_i$ to be the classically not allowed ones. Then the pair of g-many quantum action integrals
\begin{equation}
    W_{a_i}(E) = \sum_{n=0}^{\infty}\oint_{a_i}dq\frac{\rho_n}{p^{3n-1}}\hbar^n, \qquad W_{b_i}^{(n)}(E) = \sum_{n=0}^{\infty}\oint_{b_i}dq\frac{\rho_n}{p^{3n-1}}\hbar^n
\end{equation}
as well as their quantum periods
\begin{equation}
    T_{a_i}(E) = \frac{\del W_{a_i}}{\del E}, \qquad T_{b_i}(E)=\frac{\del W_{b_i}}{\del E}
\end{equation}
define a quantum lattice $\Lambda_E^Q$ with rank 2g in $\mathbb{C}^g$ \cite{Buchstaber_Bunkova_2024}. Hence one gets the Jacobi variety, a genus-g torus, $T_E^g=\mathbb{C}^g/\Lambda^Q_E$ for the curve $\e_E$ which is homeomorphic to the Riemann surface X. It is then understood that the trajectory $q_E(t)$ satisfying the QHJ is an automorphic function
\begin{equation}
    q_E(t) : \mathbb{C}^g/\Lambda_E^Q \to X
\end{equation}
parameterizing the curve $\e_E(q,p)=\e_E(q_E(t),q_E'(t))$ together with the choice of an initial condition $q_E(0)=q_0$ whose periodicity manifests itself with the invariance under the action of $PSL(2g,\mathbb{Z})$ transformation $\sigma$
\begin{equation}
    \sigma q_E(t) =q_E(t+\vec{n}\cdot\vec{T}_a + \vec{m}\cdot\vec{T}_b) = q_E(t),\quad \sigma\in\text{PSL}(2g,\mathbb{Z}), \quad (\vec{T}_a,\vec{T}_b)\in\Lambda_E^Q
\end{equation}
with $n,m$ are integer coefficients of the elements of $\sigma$ and $\vec{T}_{a}$, $\vec{T}_{b}$ are the column vectors
\begin{equation}
    \vec{T}_{\gamma_i} = \begin{pmatrix}
        T_{\gamma_1}\\
        \vdots\\
        T_{\gamma_g}
    \end{pmatrix}.
\end{equation}
Thus, any periodic path $\Gamma_E$ of the curve $\e_E$, which is the solution to the complexified classical equation of motion is associated with a quantum action $W^Q_{\Gamma}$ and a quantum period $T^Q_{\Gamma}$ who is an element of the quantum lattice $\Lambda^Q_E$ so that the following addition over the cycles is induced
\begin{equation}
    T_\Gamma^Q = n_iT_{a_i}+m_iT_{b_i} \implies \Gamma = n_ia_i + m_ib_i.
\end{equation}
The upshot is that summing over all classical periodic paths at constant energy level after the quantum canonical transformation \eqref{QCT} captures all quantum corrections and is equivalent to summing over the quantum period lattice $\Lambda_E^Q$ which is an infinite discrete sum.
%%It then follows from \eqref{ct} that the new coordinates are defined as
%\begin{align}
    %p_Q(q,\Tilde{E},\hbar) &\equiv \frac{\del S_Q(q,\Tilde{E})}{\del q}= p_{\text{classical}}+\mathcal{O}(\hbar)\\
    %\tau(q,\Tilde{E},\hbar) &\equiv \frac{\del S_Q(q,\Tilde{E})}{\del \Tilde{E}} = -t_0+\mathcal{O}(\hbar).
%\end{align}
%We have chosen our new coordinates to be the constants of motion of the classical trajectories plus their quantum corrections. For periodic trajectories, the quantum variables $\Tilde{E}$ and $\tau$ are also constants of motion. The coordinate $\tau$ is on the same footing as a time coordinate on the trajectory. For a canonical transformation, the jacobian for the path integral is just 1, i.e.,
Now, let us apply our quantum canonical transformation to the phase space path integral. The Jacobian for the functional measure is mostly 1 (after discretization), except at the initial points $q(T)=q(0)=q_0$, where we choose to keep it in the old coordinates. Keeping that in mind, we formally write,
\begin{equation}
    \mathcal{D}q\mathcal{D}p \xrightarrow{\text{C.T}} \mathcal{D}Q\mathcal{D}P = \mathcal{D}\tau\mathcal{D}\Tilde{E}.
\end{equation}
The action in the exponent transforms as (for periodic boundary conditions)
\begin{equation}
    \int_0^T dt\left(p\Dot{q}-H\right) \xrightarrow{\text{C.T}} \int_0^T dt\left(P\Dot{Q}+\frac{\del S_Q}{\del t}\right) = \int_0^{T} d\tau\tilde{E} + S_Q(q(T),\tilde{E}(T))
\end{equation}
where $S_Q$ only depends on the final old coordinate $q(T)$ and the final new momentum $\Tilde{E}(T)$,
\begin{equation}
    S_Q(q(T),\Tilde{E}(T)) = -\Tilde{E}(T)T + \int^{q(T)}_{q(0)}p_Q\left(q',\Tilde{E}(T)\right)dq'
\end{equation} 
\subsection{The Path Integral as a Discreet Sum Over Quantum Period Lattice}
 In the functional integral \eqref{resolvent}, we are performing a sum over arbitrary periodic paths in phase $(p, q)$  space. Now that 
we have transformed the phase space into the coordinates    $(\Tilde{E},  \tau)$, we can alternatively talk about  
 periodic paths in the    $(\Tilde{E},  \tau)$ space, and perform a path integral therein. Because of the quantum canonical transformation, the paths are now restricted to live in the moduli space of the curve $\e_E$. We are now in a position to take the path integral for $G(E)$,
\begin{align}
    G(E) &= \frac{i}{\hbar}\int_0^{\infty} dT \int dq_0\int_{q(0)=q(T)=q_0}\mathcal{D}[\tau]\mathcal{D}[\Tilde{E}]e^{\frac{i}{\hbar}\left(\int\Tilde{E}d\tau +ET -\Tilde{E}_fT + \int_{q_0}^{q(T)}p_Q(q',\Tilde{E}_f)dq'\right)}\end{align}
    We separate the endpoints and rearrange the $\tau_k$ integrals in order to take $\Tilde{E_k}$ integrals first
    \begin{align}
    \nonumber
    &= \frac{i}{\hbar}\int_0^{\infty} dT\int \frac{dq_0dp_0}{2\pi\hbar}\lim_{N\to \infty}\int\cdots\int\left(\prod_{k=1}^N\frac{d\tau_kd\Tilde{E}_k}{2\pi\hbar}\right)\\&\qquad\qquad\nonumber\times\exp{\frac{i}{\hbar}\left(-\sum_{k=1}^N\tau_k(\Tilde{E}_{k}-\Tilde{E}_{k-1}) - \Tilde{E}_0\tau_0 + \Tilde{E}_N\tau_{N+1} +ET -\Tilde{E}_{N}T + \int_{q_0}^{q(T)}p_Q(q',\Tilde{E}_N)dq'\right)}\\\nonumber
    &= \frac{i}{\hbar}\int_0^{\infty} dT\int \frac{dq_0dp_0}{2\pi\hbar}\lim_{N\to \infty}\int\cdots\int\left(\prod_{k=1}^{N}d\Tilde{E}_k\delta(\Tilde{E}_{k-1} - \Tilde{E}_{k})\right)\\&\qquad\qquad\nonumber\times\exp{\frac{i}{\hbar}\left( - \Tilde{E}_0\tau_0 + \Tilde{E}_N\tau_{N+1} +ET -\Tilde{E}_{N}T + \int_{q_0}^{q(T)}p_Q(q',\Tilde{E}_N)dq'\right)}
    \end{align}
    Taking the $\Tilde{E_k}$ integrals will set every $E_k$ to $E_0$. After that, we make a coordinate transformation of $p_0$ to $\Tilde{E}_0$, as well as using the periodic boundary conditions in the new coordinates $\tau_{N+1}=\tau_0$, gives
    \begin{align}
    \nonumber
    &=\frac{i}{\hbar}\int_0^{\infty} dT\int \frac{d\Tilde{E}_0}{2\pi\hbar}\int dq_0\frac{\del^2S_Q(q_0,\Tilde{E_0})}{\del q_0\del\Tilde{E}_0}\exp{\frac{i}{\hbar}\left(ET -\Tilde{E}_0 T +\int^{q(T)}_{q_0}p_Q(q',\Tilde{E}_0)dq'\right)}
     \end{align}
     It is important to mention that changing the initial point of the Abel-Jacobi map $\int_{q_0}^{q}\omega$ at a fixed $\Tilde{E}_0$ will shift the map on the torus $\mathbb{C}^{\cg}/\Lambda$, so its integration for periodic paths $q(T)=q_0$ will yield a sum over topologically distinct cycles (prime periodic orbits). This is usually overlooked in other analyses of this topic, where the endpoint dependency of the reduced action integral is ignored. Hence, the integral over $q_0$ yields
     \begin{align}
     \nonumber 
    &= \frac{i}{\hbar}\sum_{\gamma\,\in\,\text{p.p.o}}(-1)^{n_{\gamma}}\int_0^{\infty}dT_{\Gamma}\int \frac{d\tilde{E}_0}{2\pi\hbar}T_{\gamma}(\Tilde{E}_0)\exp{\frac{i}{\hbar}\left((E -\Tilde{E}_0 )T_{\Gamma} +\oint_{\Gamma}p_Q(q',\Tilde{E}_0)dq'\right)}
\end{align}
where $(-1)^{n_{\gamma}}$ is the Maslov index, achieved after integrating along the prime periodic orbit once, so for the $\gamma\in\text{p.p.o}$, $n_{\gamma} = 1$ and $T_{\gamma}$'s are the quantum periods that are elements of the quantum lattice $\Lambda^Q_{\Tilde{E}_0}$. Notice that $\gamma$ still has an implicit dependency on $T_{\Gamma}$ due to the fact that $T_{\gamma} = T_{\Gamma}/n_{\Gamma}$ where $n_{\Gamma}$ is the number of windings over the prime periodic orbit, i.e., $\Gamma=n\gamma$. We have changed the integration variable $T\to T_{\Gamma}$ to distinguish between the period of the prime periodic orbit and the total period. We now perform the $T_{\Gamma}$ integration, which will yield a positively restricted sum over the period lattice for the action, and one gets
\begin{align}
    \nonumber
    = -i\sum_{n=1}^{\infty}\sum_{\gamma\,\in\,\text{p.p.o.}}\frac{(-1)^{n}}{n}\int \frac{d\Tilde{E}_0}{2\pi}\frac{1}{(E-\Tilde{E}_0)^2}\exp{\frac{i}{\hbar}\left(n\oint_{\gamma}p_Q(q',\Tilde{E}_0)dq'\right)}
\end{align}
A residue calculation would set $\Tilde{E}_0=E$ and give the final result
\begin{equation}
    G(E) = \frac{i}{\hbar}\sum_{n_{\gamma}=1}^{\infty}\sum_{\gamma\,\in\,\text{p.p.o.}}(-1)^{n_{\gamma}}T_{\gamma}(E)\exp{\frac{i}{\hbar}\left(n\oint_{\gamma}p_Q(q,E)dq\right)}
\end{equation}
Finally, we arrive at our main result. A summary of what has been done is as follows. The quantum action $S_Q$ that satisfies the quantum Hamilton-Jacobi equation can be used to define a canonical transformation. This restricts the contributions of the paths in the path integral to be in the moduli space $\mathcal{M}_{\e}$ of the hyperelliptic curve $\e_E$. Path integration is then set to be the sum over all periodic paths at the level sets of the classical Hamiltonian. In other words, the sum is over all connected classical trajectories, including the ones with purely complex momenta. One might think that this is just a semiclassical approximation but the quantum action satisfies the QHJ equation, which is equivalent to the Schrödinger equation. Hence, all the quantum corrections are captured and related to the classical trajectories, as explained earlier. It is then understood that the spectral path integral for the resolvent is a positively restricted (implying forward propagation in time) sum over the quantum lattice $\Lambda_E^Q$ generated by the fundamental quantum reduced actions $\left\{\left(W_{a_i},W_{b_i}\right)\right\}$ and the quantum periods $\left\{\left(T_{a_i},T_{b_i}\right)\right\}$,
\begin{equation}
    G(E) = \frac{i}{\hbar}\sum_{(W_{\Gamma},T_{\Gamma})\in\Lambda_{\e}^+}\frac{(-1)^{n_{\gamma}}}{n_{\gamma}}T_{\Gamma}(E)e^{\frac{i}{\hbar}W_{\Gamma}(E)}\label{eq:latticesum}
\end{equation}
here $n_{\gamma}$ is the winding number of the prime periodic orbit $\gamma$ at each element of the lattice. One can define the prime periodic orbits as the minimal connected periodic paths that can generate all other possible periodic paths, whose periods and actions are a linear combination of the generators of the period lattice $\Lambda_{\e}$.

\section{Strange Instanton Effects%: Is the exact solution more important than its clusters?
}
\label{sec:neweffects}
The spectrum of the system is mapped to the poles of the resolvent \eqref{trace}. Since it is the derivative of the logarithm of the spectral determinant,
\begin{equation}
     -\frac{\del}{\del E}\log{D(E)} = G(E),
\end{equation}
the spectrum can be found by solving the roots of the equation,
\begin{equation}
    \implies D(E) = 0 
\end{equation}
which is the exact quantization condition for a given bounded system. It has the same equivalent form of the exact quantization conditions found in exact WKB analysis. For demonstrative examples on how to use the formula \ref{eq:latticesum} we refer to the appendix section \ref{sec:examples}. For more detailed information and examples on exact WKB analysis and quantization conditions we refer to \cite{sueishi_exact-wkb_2020}.\par
The exact quantization condition can be used both as a qualitative tool as well as a quantitative tool to learn about the dynamics of quantum mechanical systems. In this section, we use it as a quantitative tool to deduce some  strange sounding non-perturbative effects.  Trying to translate the implications of these effects to more standard instanton language teaches us some valuable lessons about instantons, multi-instantons and their role in dynamics. 

In particular, we will show that in generic multi-well systems, despite the fact that instantons are exact saddles, they do not contribute to the energy spectrum 
at leading order.  The leading NP contributions are from critical points at infinity, correlated two-events or other clusters of the instantons. 
The leading order instanton contributions in double-well and periodic potentials seems to be an exception, rather than the rule. And we would like to explain this within this section. 

 \begin{figure}[h]
    \centering
    \includegraphics[scale=0.3]{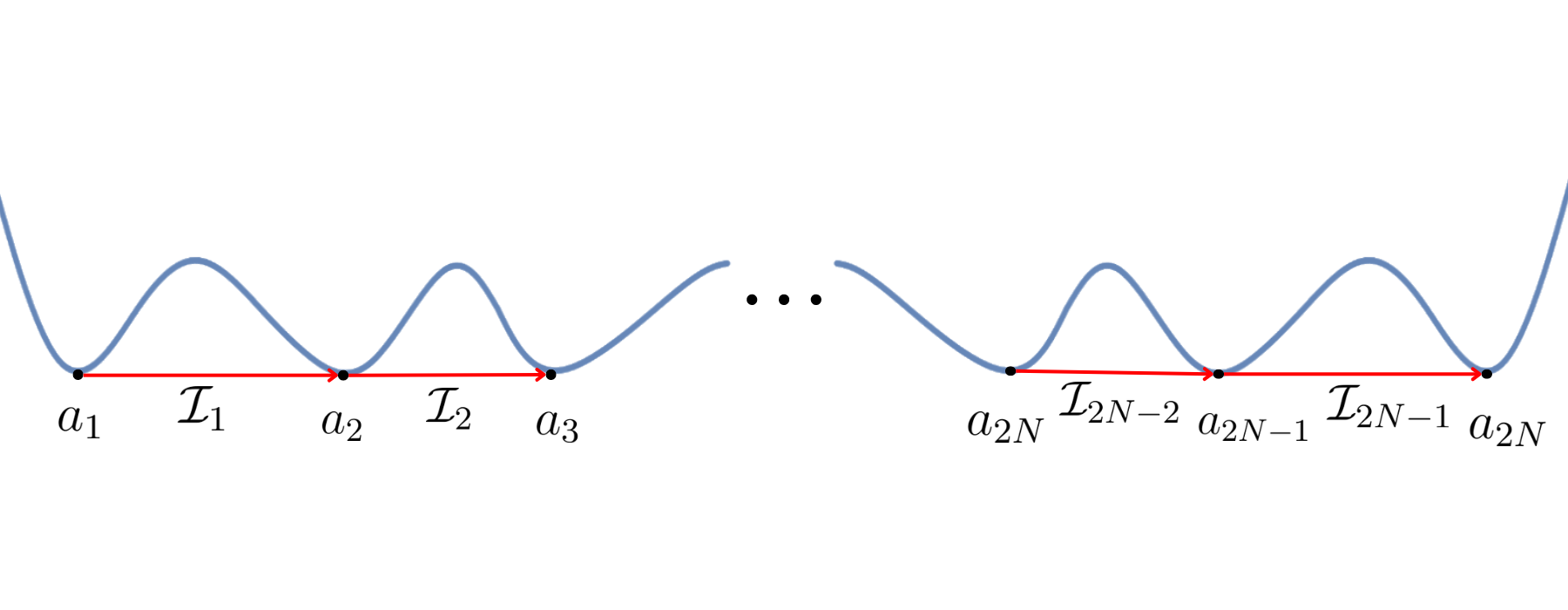}
    \vspace{-1cm}
    \caption{$\mathbb{Z}_2$ symmetric $2N$-well potential. For the generic potential of this type, consecutive wells are non-degenerate.  
    }
    \label{fig:Z2well}
\end{figure}

Consider a generic $2N$-well potential with $\mathbb{Z}_2$ reflection symmetry. 
\begin{align}
V(x)= \half \prod_{i=1}^{2N} (x-a_i)^2,  \qquad a_i= a_{2N+1-i}
\end{align} 
We assume that the  frequencies are unequal, except the  ones enforced by  $\mathbb{Z}_2$ symmetry,  
\begin{align} 
\omega_i =\omega_{2N+1-i},  \qquad    \omega_i^2 = \prod_{j \neq i} (a_j-a_i)^2  
\end{align}
Clearly, there exist instanton solutions interpolating between adjacent vacua. 
Let us enumerate them as 
\begin{align}
a_1 \xrightarrow[ \;\; I_1 \;\; ]{} a_2 \xrightarrow[ \;\;I_2 \;\;]{} a_3   \xrightarrow[\;\; I_{3} \;\; ]{}  \cdots    \xrightarrow[I_{2N-2}]{}  a_{2N-1} \xrightarrow[I_{2N-1}]{} a_{2N}
%I_{i}: \quad  a_i \ a_{2N+1-i}
\end{align} 
where $I_i$ denotes instanton interpolating from  $|a_i \rangle$ to  $|a_{i+1} \rangle$.  
We would like to understand their role in non-perturbative dynamics by using exact quantization conditions as a guiding tool. The exact quantization condition produces some results that may seem exotic from the instanton point of view.  

We would like to determine the level splittings between the perturbatively degenerate lowest states in each well, $| a_i \rangle 
 \leftrightarrow  | a_{2N+1-i} \rangle$, for $i=1, \ldots, N$. 
  By solving exact quantization conditions,  we find the leading order level splitting to be the following:
\begin{align}
\Delta E_1 & \sim  \sqrt{ B_1 B_2 \ldots  B_{2N-2} B_{2N-1} } \sim I_1 I_2 \ldots  I_{2N-2} I_{2N-1} \sim e^{-(S_1+ S_2+  \ldots   + S_{2N-1})} \cr 
\Delta E_2 &= \sqrt{  B_2 \ldots  B_{2N-2}  } \sim  I_2 \ldots  I_{2N-2}  \sim e^{-(S_2+  \ldots   + S_{2N-2})} \cr  
\ldots  \cr
\Delta E_N &= \sqrt{  B_N  } \sim   I_{N}  \sim e^{-(S_{N})} 
\label{splitting}
\end{align} 
This can also be seen at an intuitive level as follows.  The leading contribution to the transition amplitude between the perturbatively degenerate states $|a_i \rangle$ and $ |a_{2N+1-i} \rangle$ is given by: 
\begin{align}
 \langle  a_{2N+1-i}  | e^{-\beta H}  |a_i \rangle \sim e^{-(S_i + \ldots + S_{2N-i}) }
\end{align} 
which is a  a $(2N+1 -2i)$- instanton effect.  The states 
\begin{align}
|\psi_{i, \pm}\rangle  = \frac{1}{\sqrt {2}} ( |a_i \rangle  \pm |a_{2N+1-i} \rangle) 
\end{align} 
are split via an  $(2N+1 -2i)$- instanton effect. This is  a 1-instanton effect for $i=N$ and   $(2N-1)$-instanton effect for $i=1$.

% In other words, the  states   $|\psi_{1, \pm}\rangle  = ( |a_1 \rangle  \pm |a_{2N} \rangle) / \sqrt {2}$ are  split at the leading order via  an $(2N-1)$-instanton effect. More generally,  the states 
% \begin{align}
% |\psi_{i, \pm}\rangle  = \frac{1}{\sqrt {2}} ( |a_i \rangle  \pm |a_{2N+1-i} \rangle) 
% \end{align} 
% are split via an  $(2N+1 -2i)$- instanton effect. Finally, 
% the splitting between   $|\psi_{N, \pm}\rangle  = ( |a_N \rangle  \pm |a_{N+1} \rangle) / \sqrt {2}$ is a
%  $1$-instanton effect. 
 
Despite the fact that instantons $I_i$ exist as saddles, the leading splitting between   $|\psi_{i, +}\rangle  $ and $|\psi_{i, -}\rangle  $ is $(2N+1 -2i)$- instanton effect!
 Thus, we learn that except for adjacent middle minima, the level splitting is never a 1-instanton effect. In this sense, the double-well potential is an exceptional case. This said,  it is worthwhile emphasizing that  \eqref{splitting} is not generically  the leading non-perturbative effect. There are much larger NP effects but they do not lead to level splitting, rather they lead to the overall shift of the energy eigenvalue. For example, let us express various contributions to $E_{1, \pm}$. We find 
 \begin{align}
% E_{1,+}  = \half \omega_1(1+ O(\hbar))  + c_1 B_1 + c_2 B_1 B_2 + \ldots + c_{N-1} B_1 \ldots B_{N-1} 
%- \sqrt{ \prod_{i=1}^{2N-1} B_i } \cr
E_{1,+}  = \half \omega_1(1+ O(\hbar))  + c_1 B_1 + c_2 B_1 B_2 + \ldots + c_{N-1} B_1 \ldots B_{N-1} 
- \sqrt{ B_1 B_2 \ldots  B_{2N-2} B_{2N-1} } \cr
E_{1,-}  = \half \omega_1(1+ O(\hbar))  + c_1 B_1 + c_2 B_1 B_2 + \ldots + c_{N-1} B_1 \ldots B_{N-1} 
+ \sqrt{ B_1 B_2 \ldots  B_{2N-2} B_{2N-1} }
\label{splitting2}
\end{align} 
 The 2-instanton , 4-instanton, \ldots,  $(2N-2)$-instanton  
 effects are also present, but they do not lead to level splitting. The leading order level splitting comes from $(2N-1)$ instantons. 
 Eq.\eqref{splitting2} comes from the solution of exact quantization condition. How do we understand it from  the standard instanton analysis of path integral? Why do exact saddles do not contribute, but their clusters do? 
 
Instantons are solution  for the non-linear equations:\footnote{Our convention for instantons is shown in Fig.~\ref{fig:Z2well}.
They are tunneling from $a_i$ to $a_{i+1}$, and $x(\tau)$ is always an increasing function.   If  $\dot x = + \sqrt{2V}$ is the solution for $I_1$, then the $I_2$ configuration must be  the solution of  $\dot x = - \sqrt{2V}$, and this continues in this alternating manner. This alternation is due to the fact that our left hand side $\dot x$ is increasing in our convention. Yet,   $\sqrt{2V}=\prod_{j=1}^{2N}  (x-a_j)$ switches signs at all $x=a_i$.   }
 \begin{align}
 \dot x= \pm  \prod_{j=1}^{2N}  (x-a_j),  \quad x(-\infty) =a_i, \quad x(+\infty) =a_{+i+1},
 \end{align}
 It is easy to solve for the inverse function $\tau(x)$. However the function $x(\tau)$ is much harder to understand. The function $x(\tau)$ is the singular limiting case of the finite energy solution $x_E(\tau)$ and is an $2\cg = 2(2N-1)$-times periodic automorphic function (for $\cg>1$). Finding the uniformizing function $x_E(\tau)$ of the underlying Riemann surface $\mathbb{C}^{\cg}/\Lambda_E$ is one of the most challenging problems in modern mathematics.   The inverse function 
 is given by: 
  \begin{align}
\tau(x) + {c} = \log \left[ \frac{ (x -a_2)^{\frac{1}{\omega_2}} \ldots (x -a_{2N})^{\frac{1}{\omega_{2N}}} \qquad } {(x -a_1)^{\frac{1}{\omega_1}} \ldots (x -a_{2N-1})^{\frac{1}{\omega_{2N-1}}} } \right]
\label{inverse}
 \end{align}
The solutions are exact, they have finite action, and  one would naively expect a contribution to the spectrum of the form 
$e^{-S_{i}}$ from these configurations.

However,  the full instanton amplitude also includes the determinant of the  fluctuation operator around the instanton solution. The amplitude is of the form
  \begin{align}
I_{i} \sim  J_{\tau_0} \left[  \frac{  {\det}^{'}(M_i)}        {\det{(M_0)}} \right]^{-\frac{1}{2}}
 e^{-S_i}, \qquad  M_i = - \frac{d^2 }{d \tau^2} + V^{''}(q(\tau)) \Big|_{ q(\tau) = q_{{\rm cl}, i} (\tau)}
\label{Inst}
\end{align} 
 The prime indicates that the zero mode is omitted from the determinant. It must be 
 integrated over exactly, with a measure given by the Jacobian factor  $J_{\tau_0}= \sqrt{{S_i}/{2\pi}}$.      
   ${\det{(M_0)}}$ %, where    $M_0 = - \partial_{\tau}^2 + \omega_0^2$,  
      is the normalization by the
free fluctuation operator around the perturbative vacuum, present to regularize the determinant.  
 The crucial property of the generic fluctuation operator  is its asymmetry. This  generates 
 the difference with respect to  double-well and periodic potential examples, for which  $M_i$ is 
 $\mathbb Z_2$ symmetric.  The  asymmetry of the fluctuation operator  implies that  its  determinant  remains infinite even after regularization.

 \begin{figure}[th]
    \centering
    \includegraphics[scale=0.14]{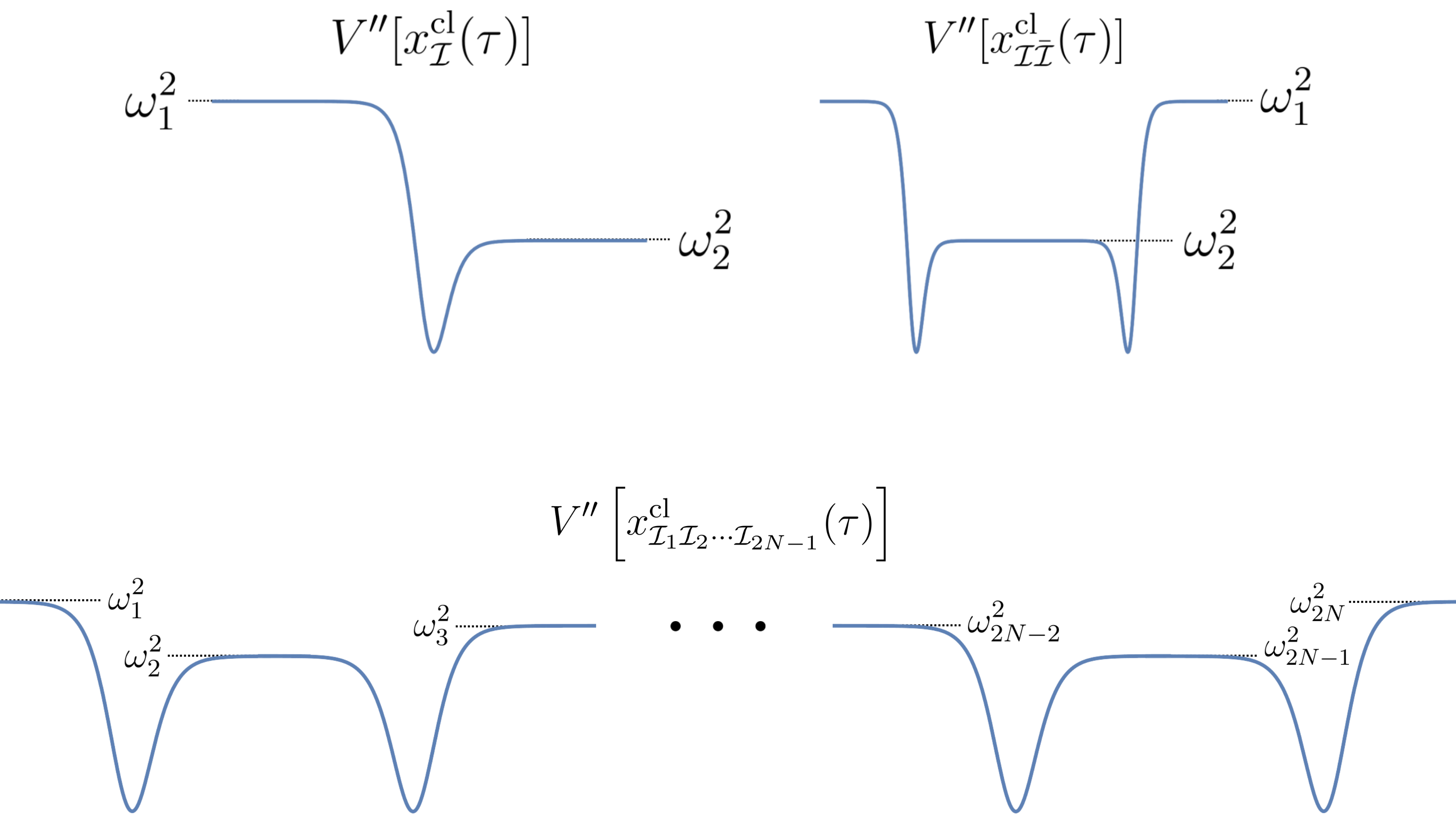}
    \vspace{-0cm}
    \caption{ Generic fluctuation operators for instantons and bions in 
    $\mathbb{Z}_2$ symmetric $2N$-well potential.  The asymmetry of fluctuation operator lead to vanishing of instanton amplitude, while the $[I_i \bar I_i]$ and  $[I_i \ldots  I_{2N-i}]$ has symmetric fluctuation operator and contribute to the energy spectrum.
    }
    \label{fig:Z2wellfluc}
\end{figure}

To determine the determinant of  fluctuation operator, we need the asymptotic profile of the instanton solution at least asymptotically,   
 As $\tau \rightarrow -\infty$, $x (\tau) \rightarrow a_i$ and $\tau \rightarrow \infty$, $x (\tau) \rightarrow a_{i+1}$. Let us write $x(\tau) = a_i + \delta_1 $ and  $x(\tau) = a_{i+1} + \delta_2 $ at two asymptotes. It is easy to determine both $\delta$, as well 
 as $V^{''}(a_i), V^{''}(a_{i+1})$. We find 
 \begin{align} 
 x_i (\tau)  &=  \left \{  \begin{array}{ll}  a_i+ e^{+ \omega_i \tau} & \quad  \tau \rightarrow -\infty \cr
a_{i+1}- e^{- \omega_{i+1} \tau} & \quad  \tau \rightarrow +\infty 
 \end{array} \right. \cr 
V^{''}(q_{{\rm cl}, i} (\tau) )  &=  \left \{  \begin{array}{ll}   \omega_i^2   & \quad  \tau \rightarrow -\infty \cr
 \omega_{i+1}^2  & \quad  \tau \rightarrow +\infty 
 \end{array} \right.
 % \Psi_0 (\tau) =  \dot   x_i (\tau) &=  \left \{  \begin{array}{ll} \omega_i  e^{+ \omega_i \tau} & \quad  \tau \rightarrow -\infty \cr
 % \omega_{i+1} e^{- \omega_{i+1} \tau} & \quad  \tau \rightarrow +\infty 
 % \end{array} \right. 
  \end{align}
  as we can  derive from the exact inverse-solution \eqref{inverse}. We do not need the full form of the fluctuation operator to show the vanishing of the instanton, and non-vanishing of bions, and certain other clusters. 
 The crucial point here, compared to the instantons in double-well potential, is the generic asymmetry of 
   $V^{''}(q_{{\rm cl}, i} (\tau))$ as $\tau \rightarrow \pm \infty$, shown in Fig.\ref{fig:Z2wellfluc}.  
Because of this, the  regularized determinant still remains infinite: 
 \begin{align}
 \left[  \frac{  {\det}^{'}(M_i)}        {\det{(M_0)}} \right]^{-\frac{1}{2}} = \lim_{\beta \rightarrow \infty} e^{ -\frac{\beta}{2} |\omega_i - \omega_{i+1}|}  =0
 \label{fluc-det}
\end{align} 
Hence, the instanton amplitude in this system is ironically zero despite the fact that instanton  configuration is finite action. 
\begin{align}
    I_i =0, \; \bar I_i =0,  \qquad i=1, 2N-1
\end{align}
 
 On the other hand, if we consider a bion $B_i = [I_i \bar I_i]$,  which is not an exact solution due to interactions between the instantons,  has a symmetric profile:  
 $x_{B,i} (\tau) \rightarrow a_i +  e^{+ \omega_i \tau}$   as  $\tau \rightarrow - \infty$,  $x_{B,i} (\tau) \rightarrow a_i +  e^{- \omega_i \tau}$   as  $\tau \rightarrow + \infty$, as shown in Fig.\ref{fig:Z2wellfluc}.  
 The fluctuation operator is symmetric and the prefactor $ \left[   {\det}^{'}(M_i)   /   \det(M_0) \right]^{-1/2}$ is finite.

 Similarly, for the correlated events  $[I_i  I_{i+1} \ldots I_{2N+1-i}], \;  i=1, \ldots, N$, the fluctuation operators are symmetric and   finite as well, see Fig.\ref{fig:Z2wellfluc}.    As a result, this multi-instanton  leads to the transition amplitude  between  the states $|a_i \rangle$    and  $|a_{2N+1-i} \rangle$:
\begin{align}
    \langle a_{2N+1-i}  | e^{-\beta H } |a_i \rangle   \sim \exp{  \Big[ - \sum_{j=i}^{2N-i} S_j \Big]
     }, \qquad i=1, \ldots, N
\end{align}
 but far more suppressed than the instanton effect.
 In our system, this is the leading configuration that can lead to level splittings!

 In the standard configuration space path integral, we conclude that only saddles with finite determinants for their fluctuation operators (after regularization) contribute to the spectrum. In the present case, this amounts to configurations with symmetric fluctuation operators.

\section{Conclusion}
In this work, we proposed a reformulation of the path integral motivated by the classical Hamilton-Jacobi theory. 
Recall that in the usual Feynman path integral  $Z(T) =\tr (e^{-i HT})$,  one considers a sum over all periodic trajectories satisfying the boundary conditions at a given {\it fixed} time interval $T$,   while the energy $E$ of such paths can take any value.   
On the other hand, in the Hamilton-Jacobi formalism,  we can describe paths without saying anything about how the motion occurs in time. We essentially wanted to achieve this in path integral via semi-classics.  This, of course, requires working with the Fourier transform of path integral,   $\tilde Z(E)$. Now,  $E$ is kept fixed, and $T$ can take arbitrarily large values.     The path integral is turned into a discrete sum over the classical periodic orbits in terms of Hamilton's characteristic function at fixed energy.

The periodic orbits that enter the story are not only the classically allowed  (perturbative) orbits of potential problems with $V(q)$. Classically forbidden periodic orbits (non-perturbative)  also enter our description.  It is worthwhile recalling that classically forbidden 
periodic orbits of  $V(q)$ are same as classically allowed orbits of $-V(q)$.
  The path integration instructs us to sum over all vanishing cycles, at  energy level $E$,
 $  (\gamma_i, \gamma_{{d}, j})  \in \Gamma$.  They are on a similar footing in the path integral perspective, except that the former is pure phase 
$ |e^{\frac{i}{\hbar}W_{\gamma_i}(E, \hbar)}|  =1$ and the latter is exponentially suppressed,   $ | e^{\frac{i}{\hbar}W_{\gamma_{ {d}, i}} (E, \hbar)} |   <1 $, related to tunneling. 
 These two factors are called the Voros symbols in the exact WKB formalism, $A_i, B_i$ \cite{Voros1983}. 
The semiclassical expansion is done around each topologically distinct cycle on the constant energy slice of the phase space, which is generically 
a non-degenerate  torus of genus-$g=N-1$.  Each topologically distinct cycle  corresponds to fundamental periods of the tori thereof.

Although not sufficiently appreciated, in a certain sense,  the {\it old quantum theory} of the pre-Schr\"odinger era, underwent a silent and slow revolution in the last decades starting with the pioneering work of  Gutzwiller  \cite{gutzwiller_periodic_1971}, in which he re-posed the question 
{\it ``What is the relation between the periodic orbits in the classical system and the energy levels of the corresponding quantum system?"}, and provided a partial answer through his trace formula of the resolvent.    Later studies, on 
generalized Bohr-Sommerfeld quantization  \cite{Zinn-Justin:2004vcw, Zinn-Justin:2004qzw}, exact WKB  
\cite{dillinger_resurgence_1993,DDP2, balian_discrepancies_1978,Voros1983, Silverstone, 
 aoki1995algebraic, AKT1, sueishi_exact-wkb_2020}, uniform WKB \cite{Dunne:2013ada, Dunne:2014bca, Alvarez3}
 are some of the works addressing this general problem from different perspective.   Now, we start to see more directly  that 
 path integrals in phase space, when performed using ideas from the  old  Hamilton-Jacobi theory in classical mechanics, produce 
the spectral path integral $\tilde Z(E)$, which is equivalent to resolvent and simply related to spectral determinant $D(E)$. 
The vanishing  of the determinant gives  the generalized and exact version of the Bohr-Sommerfeld quantization conditions.  This is the sense in which classical paths and quantum spectrum are connected. 
 
\noindent
{\bf What did we gain?} 
In the standard implementation of the path integral, we write 
\begin{equation}
Z(T)= \int \mathcal{D}[q(t)]\mathcal{D}[p(t)] \; e^{\frac{i}{\hbar}\left( \int_0^T (p \dot q -  H) dt  \right)}.
\label{PSPI}
\end{equation}
In the integration, $p(t)$ and $q(t)$ are independent  (real) variables to begin with.  However, once we start talking about semi-classics, we first pass to the complexification of these generalized coordinates.

In semi-classics, we must first find the critical points. These are given by the  (real and complex) solutions of the 
 { \it complexified }  versions of Hamilton's equations: 
\begin{align}
\frac{dq}{dt} & = \frac{\partial H}{\partial p}, \cr
\frac{dp}{dt} & = - \frac{\partial H}{\partial q}, 
\label{Hamilton}
\end{align}   
where $H(p, q)$ is viewed as a holomorphic function of $p$ and $q$.
The saddle points in the phase space formulation are periodic solutions of  Hamilton's equation. Note that the dimension of the phase space is doubled.  However, this does not imply a doubling of the number of degrees of freedom. A restriction that reduces the dimension  to appropriate middle-dimensional space 
enters through the gradient flow equations \cite{Witten:2010zr}.

It is trivial to realize that both real and complex solutions exist, after all this is a simple potential problem in classical mechanics.   However,  for genus $g \geq 2$ systems,
it is hard to write down {\it explicit} solutions as a function of time, let alone the structure of the determinant of the fluctuation operator. 
For genus $g=1$, exact solutions are just doubly-periodic complex functions, e.g., related to perturbative fluctuations and instantons.  

By using Hamilton-Jacobi canonical formalism, and working with spectral path integral $\tilde Z(E)$, we essentially bypass these difficulties.  Instead of describing the periodic orbits by their explicit functions $q(t)$, we describe them with their associated reduced actions $W_{\gamma_i}(E,\hbar)$ and periods $T_i(E,\hbar)$ as a function of energy.  Then, the spectral path integral turns into a discrete sum over P and NP periodic orbits, associated with vanishing cycles. Ultimately, these sums can be done exactly in terms of  Voros symbols, $A_i(E, \hbar)$ and $B_i (E, \hbar)$, as functions of the reduced actions $W_{\gamma}(E, \hbar)$.   The determination of $W_{\gamma}(E, \hbar)$'s are not trivial, but doable.

The spectral path integral brings out a promising formalism for the quantization of solitons in (1+1)-dimensional field theories. The previous work employing WKB quantization to solitons involves the usual path integral calculation methods and the result is known as the DHN (Dashen, Hasslacher, Neveu) formula \cite{DHN1,DHN2,rajaraman_non-perturbative_1975}. The spectral path integral formalism can give further intuition and ease of calculation for the spectrum of such theories.

\acknowledgments
We thank  Syo Kamata, Naohisa Sueishi, Can Kozcaz, Mendel Nguyen for useful discussions. The work is supported by U.S. Department of Energy, Office 
of Science, Office of Nuclear Physics under Award Number 
DE-FG02-03ER41260. 

\pagebreak
\appendix
\section{Appendix}
\subsection{Classical Mechanics in Terms of Conserved Quantities and Dual Classical Solutions}
We see that in this new formulation of the path integral with the use of 
quantum Hamilton-Jacobi theory (as in the exact WKB analysis), the summations are done over multiples of the classical periodic paths and their quantum corrections. The curious thing is that the path integral not only captures the classically allowed periodic paths but also the contributions around the dual classical solutions, which have purely imaginary actions. These are the instanton-like solutions coming from the inverted potential. To interpret these solutions in the reduced action formalism, let us start by defining the corresponding dual conjugate variables.

Let us assume that we have a stable potential $V(q)$ with $N+1$ local degenerate real minima at $V_{\text{min}}=0$ and $N$ local degenerate real maxima at $V_{\text{max}}=E_{\text{top}}$ with each extremum allowed to have different frequencies $\omega_k$. Although the following arguments will apply to non-degenerate cases as well, for the sake of simplicity, we will stick with the 
classically degenerate case. For a periodic motion at a given energy level $E$, one has $N+1$-many classically allowed periodic orbits around their minima with conserved reduced actions,
\begin{equation}
    W^{(i)}(E) = \oint_{\gamma_{i}}pdq,\qquad i= 1,2,\dots,N+1
\end{equation}
where $\gamma_i$ is the integration cycle in the complex $q$-plane, encircling the turning points $q_{i}$ and $q_{i+1}$ defined as the elements of the ordered set of solutions to the algebraic equation
\begin{equation}
    E = V(q)
\end{equation}
and classical momentum is defined by the curve
\begin{equation}
    p^2 = 2\left(E-V(q)\right).
\end{equation}
We define the action variables as
\begin{equation}
    I^{(i)}(E) = \frac{1}{2\pi}W^{(i)}(E).
\end{equation}
Observe that Hamilton's characteristic function $W(q)= \int^q dq' \sqrt{2(E-V(q'))}$ works as a type-II generating function for the canonical transformation to the action-angle variables $(q,p)\to (\phi,I)$ 
\begin{equation}
    \frac{\del W(q,E)}{\del q} = p,\qquad \frac{\del W(q,E)}{\del I^{(i)}} = \phi^{(i)} 
\end{equation}
and we wish to treat everything in terms of the new coordinates. The action variables are constants of motion, then Hamilton's equations of motion with new Hamiltonian $\Tilde{H}$ for each $I^{(i)}$ becomes
\begin{equation}
    \frac{d I^{(i)}}{dt} = \frac{\del \Tilde{H}}{\del \phi^{(i)}} = 0.
\end{equation} 
This implies that the new Hamiltonian $\Tilde{H} = \Tilde{H}\left(I^{(i)}\right)$ only depends on the action variables.  Then Hamilton's equation of motion for each angle-variable is
\begin{equation}
    \frac{d \phi^{(i)}}{dt} = \frac{\del \Tilde{H}}{\del I^{(i)}} = \omega^{(i)}[I^{(i)}] \implies \phi^{(i)}(t) = \omega^{(i)}t+\phi^{(i)}_0 
\end{equation}
The corresponding periods of each path are
\begin{equation}
    T^{(i)}(E) = \frac{\del}{\del E}\oint_{\gamma_{i}}pdq
\end{equation}
then the change in each angle variable for a periodic path about the corresponding well is
\begin{equation}
    \Delta\phi^{(i)} = \omega^{(i)}T^{(i)}.
\end{equation}
Observe that for a periodic path about the $i$'th minimum
\begin{align}
    \Delta\phi^{(i)} =  \oint_{\gamma_i}\frac{\del\phi^{(i)}}{\del q}dq = \oint_{\gamma_i}\frac{\del^2 W(q)}{\del q\del I^{(i)}}dq = \frac{d}{d I^{(i)}}\oint_{\gamma_i}pdq =2\pi \frac{d I^{(i)}}{d I^{(i)}} = 2\pi
\end{align}
Thus, we get that each $\omega^{(i)}$ is
\begin{equation}
    \omega^{(i)} = \frac{2\pi}{T^{(i)}}
\end{equation}

 However, this is not the complete set of conserved quantities in the system. For the quantum mechanical system, we know that the classically forbidden periodic solutions also contribute to the system's spectrum which describes the effect of tunneling. We also observe this contribution in the previously mentioned path integral description, where the reduced action captures the classically forbidden regions in phase space. To find these solutions, the usual procedure involves going into Euclidean time by a Wick rotation $t\to-it$.
In the reduced action formulation, this can be achieved by defining a dual-energy
\begin{equation}
    E_D \equiv E_{\text{top}}-E
\end{equation}
where $E_{\text{top}}$ is the local maximum of the potential. This way, observe that the classical momentum becomes purely imaginary for $0<E_D<E_{\text{top}}$,
\begin{align}
    p = \pm\sqrt{2(E_{\text{top}}-E_D-V(q))} &= \pm i\sqrt{2\left(E_D-\left(E_{\text{top}}+V(q)\right)\right)} \equiv \pm i\sqrt{2\left(E_D-V_D(q)\right)},\\
   \implies p&\equiv ip_D.
\end{align}
where the dual potential $V_D(q)\equiv -\left(E_{\text{top}}+V(q)\right)$ is just the inverted potential whose minima are shifted to zero. The dual reduced actions for the dual periodic paths and the dual action variables are defined as
\begin{equation}
    W_D^{(k)}(E_D) = i\oint_{\gamma_{d,k}}p_Ddq = 2\pi i I_D^{(k)}, \qquad k=1,2,\cdots, N
\end{equation}
where, again, the dual cycles $\gamma_{d,k}$ encircle the turning points $q_k$ and $q_{k+1}$ with $k$ even. The indices $i$ will always run over the number of minima of $V$ whereas $k$ will always run over the number of maxima of $V$. A similar analysis is done with the action
angle variables defined as
\begin{equation}
     \frac{\del W(q,E)}{\del q} = p,\qquad \frac{\del W(q,E)}{\del I_D^{(k)}} = \phi_D^{(k)} 
\end{equation}
and the corresponding period of each motion around the maximum $k$ of $V$ is defined as
\begin{equation}
    T^{(k)}\equiv i T^{(k)}_D(E_D)=i\frac{\del}{\del E_D}\oint_{\gamma_{d,k}}p_Ddq
\end{equation}
which is purely imaginary.
Using the equations of motion, again, we get
\begin{equation}
    \phi_D^{(k)}(t) = \omega^{(k)}_Dt + \phi_{D,0}
\end{equation}
with
\begin{equation}
    \omega_D^{(k)} = -\frac{2\pi i}{T_D^{(k)}}
\end{equation}
which makes it apparent for the reason of the Wick rotation $t \to -it$. The "frequencies" are another set of constants of the motion
\begin{equation}
    \omega^{(i)}(E) = \frac{2\pi}{T^{(i)}(E)}, \qquad \omega_D^{(k)}(E_D) = -\frac{2\pi i}{T_D^{(k)}(E_D)}. 
\end{equation}
\begin{figure}[hbt]
    \centering
    \includegraphics[scale=0.3]{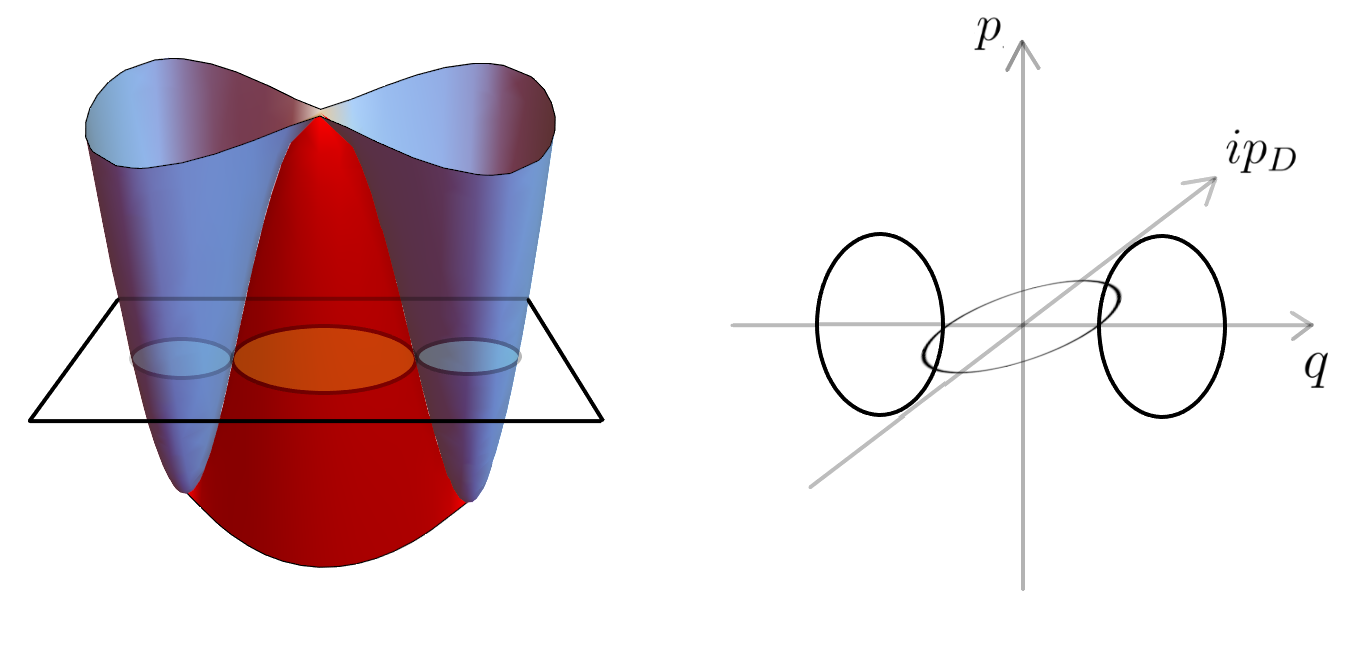}
    \caption{(Left) Elliptic surfaces traced by the periodic paths in the $(q,p,E)$-space for a double well potential. The blue one is the surface traced by the classical paths with real actions, and the red one is the surface traced by the dual classical paths with imaginary actions. The surfaces are cut by a constant energy level where the intersections are elliptic curves in phase space describing the classical motion. They are embedded in the same space for visual convenience. (Right) The loops in the complex $(q,p,p_D)$-phase space at a given constant energy slice where $p_D$ is the imaginary dual momentum.}
    \label{fig:phasespace}
\end{figure}

We see that the path integral has nonperturbative contributions coming from the real periodic paths in the classically forbidden regions with purely imaginary actions and purely imaginary periods. Notice that the motion itself $q(t)$ is still real for both classical and dual paths. Consequently, to understand the whole picture of the loops traced by the periodic orbits in the phase space $(p,q)$ at a given energy $E$, one needs to complexify the momentum in the phase as depicted in Fig.[\ref{fig:phasespace}]. The advantage of our formulation of the path integral using the quantum Hamilton-Jacobi formalism is that one does not need the explicit form of the periodic solutions $q(t)$. We only need their existence and in integrable systems where the total energy is conserved, this is always guaranteed. The classical paths are defined by the constant energy slices of the Hamiltonian seen as a height map on the phase space $H(p,q)$.

\section{Building The Summation}\label{sec:examples}

\subsection{Double Well}
To understand how the path integral summation should be carried out with systems that have instantons or bions,  we first use a simple and important example, symmetric and asymmetric double well system.  First, let us assume that the frequency of each well is different, namely, $V''(a_1) = \omega_1^2$ and $V''(a_2) = \omega_2^2$, with $a_1$, and $a_2$ being the minima of the potential. There are two perturbative $\gamma_{1,2}$ vanishing cycles and one nonperturbative $\gamma_d$ vanishing cycle. We define the  exponential of the quantum version of Hamilton's characteristic function over periodic cycles, called  Voros symbols in exact WKB as
\begin{equation}
    A_1 = e^{\frac{i}{\hbar}W_{\gamma_1}(E)}, \qquad A_2 = e^{\frac{i}{\hbar}W_{\gamma_2}(E)}, \qquad B= e^{\frac{i}{\hbar}W_{\gamma_d}(E)}
\end{equation}

\begin{figure}[ht!]
    \centering
    \includegraphics[scale=0.3]{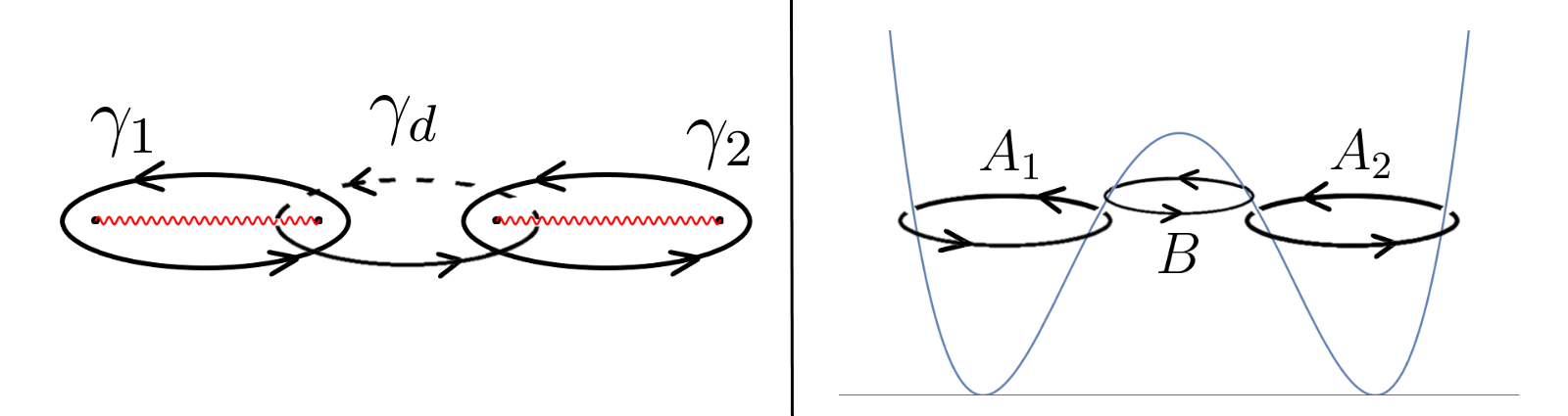}
    \caption{(Left) Integration cycles encircling the turning points on the complex $q$-plane at a nonsingular level set of $E$ for the double well system. $\gamma$ is the vanishing cycle and $\gamma_d$ is the dual cycle. (Right) Double well potential with its minimal orbits representing their corresponding Voros symbols.}
    \label{fig:DW}
\end{figure}
For now, we treat the quantum-reduced actions $W$ to be analytic functions of $E$ in the region $0 < E< V_{\text{max}}$, where $V_{\text{max}}$ is the local maximum (barrier top). The cycles $\gamma$ are defined by analytically continuing in the complex $q$-plane with $\arg\hbar<0$, encircling the classical turning points of the level set $E = p^2/2 + V(q)$. Here $\gamma_{1,2}$ cycles vanish at $E=0$ and $\gamma_d$ vanishes at $E=V_{\text{max}}$ when the encircled turning points coalesce. In the phase space of classical mechanics, the level set $E=V_{\text{max}}$ defines the separatrix.

One may think of $W(E)$'s as the Borel resummed version of their asymptotic series in $\hbar$ as in \eqref{Red-ac}. This brings about a Stokes phenomenon and an imaginary ambiguity in the analytical continuation between different choices of $\hbar>0$ and $\hbar<0$. Two choices are related by the monodromy properties of the moduli space via the transformation $E\to E'=Ee^{\pm 2\pi i}$. This amounts to going to a proper sheet and mapping one choice to the other. Also, we'll see that it maps one spectral determinant when $\hbar<0$ to the other with $\hbar >0$. Without loss of generality, we will work with $\arg\hbar<0$. This will only change the factor in front of the nonperturbative transmonomial $B$. One can show that the imaginary ambiguity cancellation occurs after the medianization of the Voros symbols but we won't pursue that direction \cite{dillinger_resurgence_1993, Aniceto:2013fka, sueishi_exact-wkb_2020}.

It is straightforward to identify the perturbative paths, they are just integer multiples of the topologically distinct cycles $\gamma_i$. Hence any other perturbative trajectory can be generated by the transmonomial
\begin{equation}
    A_i = e^{\frac{i}{\hbar}W_{\gamma_i}(E)}, \qquad i = 1, 2
\end{equation}
which gives the perturbative part of the spectral path integral for $-\log D(E)$,
\begin{align}
    -\log D_{\text{p}}(E) =& \sum_{n=1}^{\infty}\frac{(-1)^n}{n}A_1^n + \sum_{n=1}^{\infty}\frac{(-1)^n}{n}A_2^n=-\log\left[\left(1+A_1\right)\left(1+A_2\right)\right]\\\nonumber
    &\implies D_{\text{p}}(E) = (1+A_1)(1+A_2).
\end{align}

Let us now try to identify the nonperturbative transmonomial $\Phi_{\text{np}}$ by considering periodic paths that exhibit tunneling. One may be tempted to think that a similar summation would do the job. That is, any nonperturbative path is generated as an integer multiple of $\gamma_d$. \begin{equation}
    -\log D_{\text{np}}(E) = \sum_{n=1}^{\infty}\frac{(-1)^n}{n}\Phi_{\text{np}}^n \overset{?}{=} \sum_{n=1}^{\infty}\frac{(-1)^n}{n}B^n
\end{equation} However, this is incorrect. The particle can tunnel through from the left well to the right one and oscillate there $m$-times, then tunnel back. This is also a distinct periodic path for each $m$. Therefore, in general, we need to consider {\it all} possible oscillations in each well together with tunneling from left to right and vice versa. Then the path integration (summation) should be over all these paths. Ultimately, their combinations will appear in the powers of $n$ describing all possible $n$-periodic tunneling events with their binomial weights. This might sound cumbersome,   but you'll see that they repackage themselves quite nicely. Consider the following summation describing the transmonomial with all possible paths with a single periodic tunneling event.
\begin{equation}
    \Phi_{\text{np}} = \sum_{m_1,m_2=0}^{\infty}(-1)^{m_1 + m_2}BA_1^{m_1} A_2^{-m_2}
\end{equation}
Here, notice the minus sign in the exponent of $A_2$. This is because, after one tunneling event from left to right, we change Riemann sheets by going through the branch cut. We go back to the same sheet after tunneling to the left, whence no minus sign for $A_1$. We formally carry out this summation over $m_1,m_2$'s
\begin{equation}
    \Phi_{\text{np}} = \frac{B}{(1+A_1)(1+A_2^{-1})}.\label{nonpertubative}
\end{equation}
It is now easy to write down the nonpertubative spectral path integral for the determinant,
\begin{align}
        -\log D_{\text{np}}(E) &= \sum_{n=1}^{\infty}\frac{(-1)^n}{n}\Phi_{\text{np}}^n=-\log\left[1+\frac{B}{(1+A_1)(1+A_2^{-1})}\right]\\\nonumber
    &\implies D_{\text{np}}(E) = 1+\frac{B}{(1+A_1)(1+A_2^{-1})}.
\end{align}
Lastly, we arrive at the quantization condition $D(E) = 0$, from the fact that
\begin{equation}
    \log D(E) = \log D_{\text{p}}(E) +\log D_{\text{np}}(E) = \log\left[D_{\text{p}}(E)D_{\text{np}}(E)\right]
\end{equation}
one gets the exact quantization condition to be
\begin{equation}
    D(E) = (1+A_1)(1+A_2)\left(1+\frac{B}{(1+A_1)(1+A_2^{-1})}\right) = 0. \label{dwquantization}
\end{equation}
This is the same result as one would get from the WKB analysis by demanding 
normalizability of the WKB wave function. 

\begin{figure}
    \centering
    \includegraphics[scale=0.25]{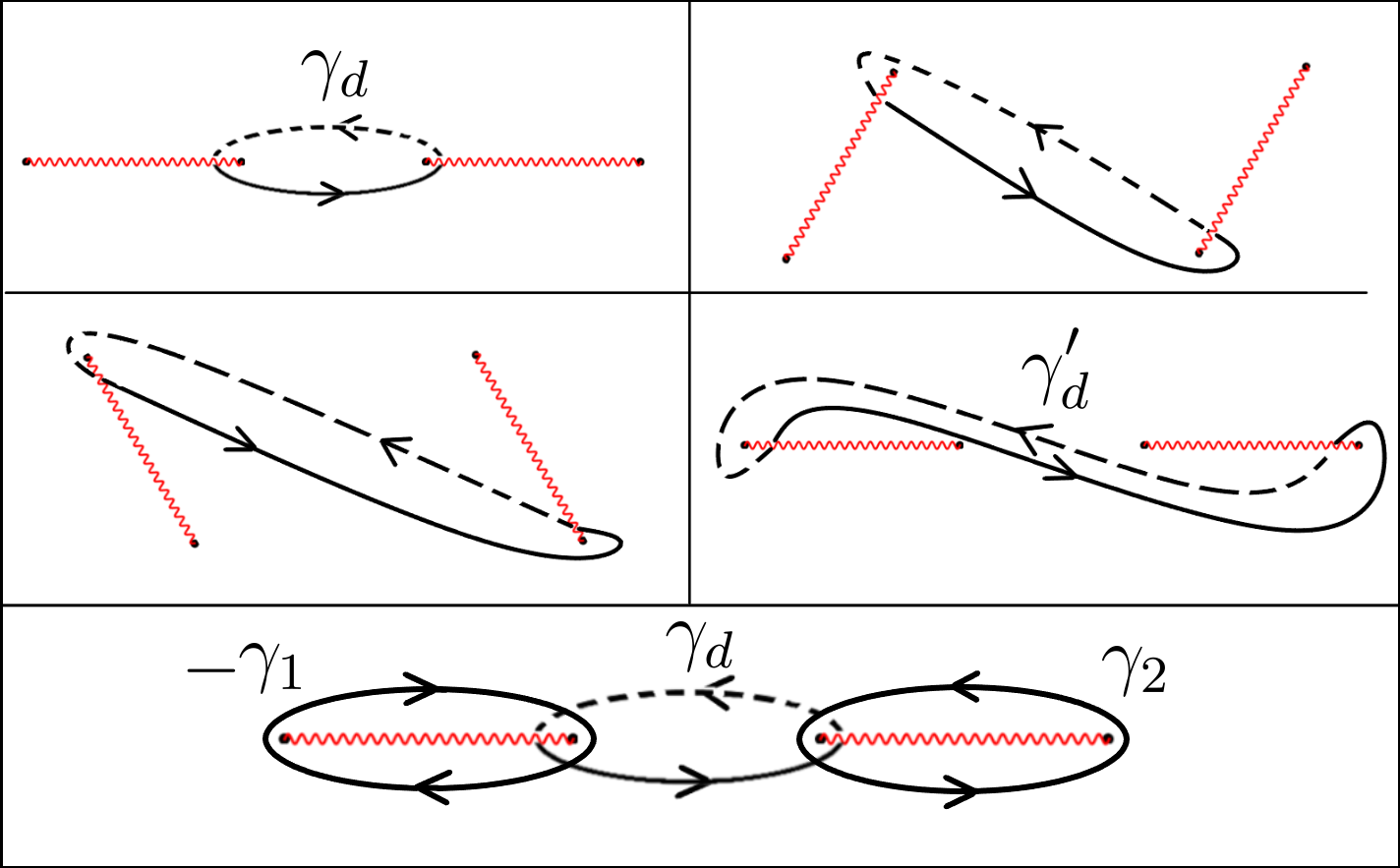}
    \caption{Action of the monodromy transformation $E\to E'=Ee^{i\theta}$. (Top Left) The branch cuts and the dual cycle $\gamma_d$ at $\theta =0$, (Top Right) at $\theta = \frac{2\pi}{3}$, (Bottom Left) at $\theta = \frac{4\pi}{3}$, (Bottom Right) at $\theta=2\pi$. (Bottom) Corresponding decomposition of the dual cycle resulting from the monodromy. It is easy to see this decomposition by blowing up the line segments to infinity after $2\pi$ rotation.}
    \label{fig:PL}
\end{figure}

Had we started with $\arg\hbar>0$, one had to choose $A_1^{-1}$ in the transmonomial \eqref{nonpertubative} instead, and would end up with
\begin{equation}
    D(E) = (1+A_1)(1+A_2)\left(1+\frac{B}{(1+A_1^{-1})(1+A_2)}\right) = 0.
\end{equation}
This is a mere choice of our definitions of the cycles in the first sheet. The two cases are related to each other via the action of the monodromy around $E=0$. To see the effect of the monodromy of $E$, consider the transformation $E\to Ee^{\pm 2\pi i}$. The quantum momentum function $p_Q(E,\hbar)$ is invariant under this transformation but the dual (co)vanishing cycle $\gamma_d$ is not. The dual vanishing cycle transforms according to the Picard-Lefschetz formula 
\begin{equation}
    \gamma_d' = \gamma_d \pm \gamma_1 \mp \gamma_2
\end{equation}
where the perturbative vanishing cycles $\gamma$ are invariant under the action of the monodromy transformation (Fig.[\ref{fig:PL}]). 
Thus, nonperturbative transmonomial, defined for $\arg\hbar<0$, transforms as
\begin{equation}
    \Phi_{\text{np}} = \frac{B}{(1+A_1)(1+A_2^{-1})} \to \Phi'_{\text{np}} =\frac{BA_1A_2^{-1}}{(1+A_1)(1+A_2^{-1})} = \frac{B}{(1+A_1^{-1})(1+A_2)}
\end{equation}
which is the same transmonomial defined for $\arg\hbar>0$.

We see that the nonperturbative transmonomial $B$ comes dressed up with the spectral determinants of the perturbative orbits alternating in separate sheets connected via tunneling.

For a symmetric double-well potential, the quantization condition becomes 
\begin{equation}
    D(E) = D_{\rm p}(E) D_{\rm np}(E)= (1+A)(1+A^{-1})\left(1+\frac{B}{(1+A)^2}\right) = 0
\end{equation}

\noindent
{\bf Application:} Apart from being a quantitatively excellent tool, the quantization condition is also  a  qualitatively useful  tool.   The leading order non-perturbative contribution to symmetric 
and asymmetric (classically degenerate) double well potentials are of different nature \cite{Miller}, as it can be seen by solving quantization conditions. One finds, the leading non-perturbative contribution to the energy spectrum as: 
\begin{equation}
    \Delta E_{\text{np}} \sim  \left\{ \begin{array}{ll} 
    e^{-S_I/\hbar}  \quad {\rm level \;  splitting},  &  \qquad {\rm symmetric \;  DW} \cr 
    e^{-2 S_I/\hbar}  \quad {\rm shift},  & \qquad {\rm asymmetric \; DW }
    \end{array} \right.
\end{equation}
For a symmetric double-well potential,  it is well-known that  the leading non-perturbative effect is level splitting, of order  $e^{- S_I}$,   which is due to an instanton.  
For an asymmetric, and classically degenerate double-well potential, with  $\omega_1<\omega_2$,  
despite the fact that instanton is a finite action saddle, the instanton contribution vanishes.  
As explained in detail in \S.\ref{sec:neweffects}, the fluctuation determinant is infinite and this renders the instanton contribution zero. 
The leading order non-perturbative contribution to vacuum energy is of order 
  $e^{-2 S_I}$, the bion (or correlated instanton-anti-instanton effect). The determinant of fluctuation operator for this 2-event is finite, and hence it contributes to the spectrum. 
  For related subtle non-perturbative phenomena and detailed explanations, see \S.\ref{sec:neweffects}.

\subsection{Quantum Mechanics on $S^1$}
Let us now apply our knowledge to quantum mechanical systems with periodic potentials $V(q)$ obeying
\begin{equation}
    V(q+a) = V(q).
\end{equation}
By Bloch's theorem, the wavefunction of a system with periodic potential attains the form
\begin{equation}
    \psi(q) = e^{ikq}u_k(q)
\end{equation}
with $u_k(q+a) = u_k(q)$, so that it satisfies 
\begin{equation}
    \psi(q+a) = e^{ika}\psi(q).
\end{equation}
The spectra of the system consist of bands whose states are labeled by the continuous Bloch momenta $ka\in [-\pi,\pi]$. Upon gauging the $\mathbb{Z}$ translation symmetry, we make the physical identification of $q+Na = q$ with $N\in\mathbb{Z}$. Hence, the space is compactified to a circle, that is, $q\in S^1$. Therefore, we can introduce a theta-angle, $\theta\equiv ka$ such that one has the relationship of the wavefunctions
\begin{equation}
    \psi(q+2\pi)=e^{i\theta}\psi(q)
\end{equation}
where we have set the lattice spacing $a=1$.
We need to account for this fact in our formulation of the spectral determinant.

\begin{figure}[h]
    \centering
    \includegraphics[scale=0.22]{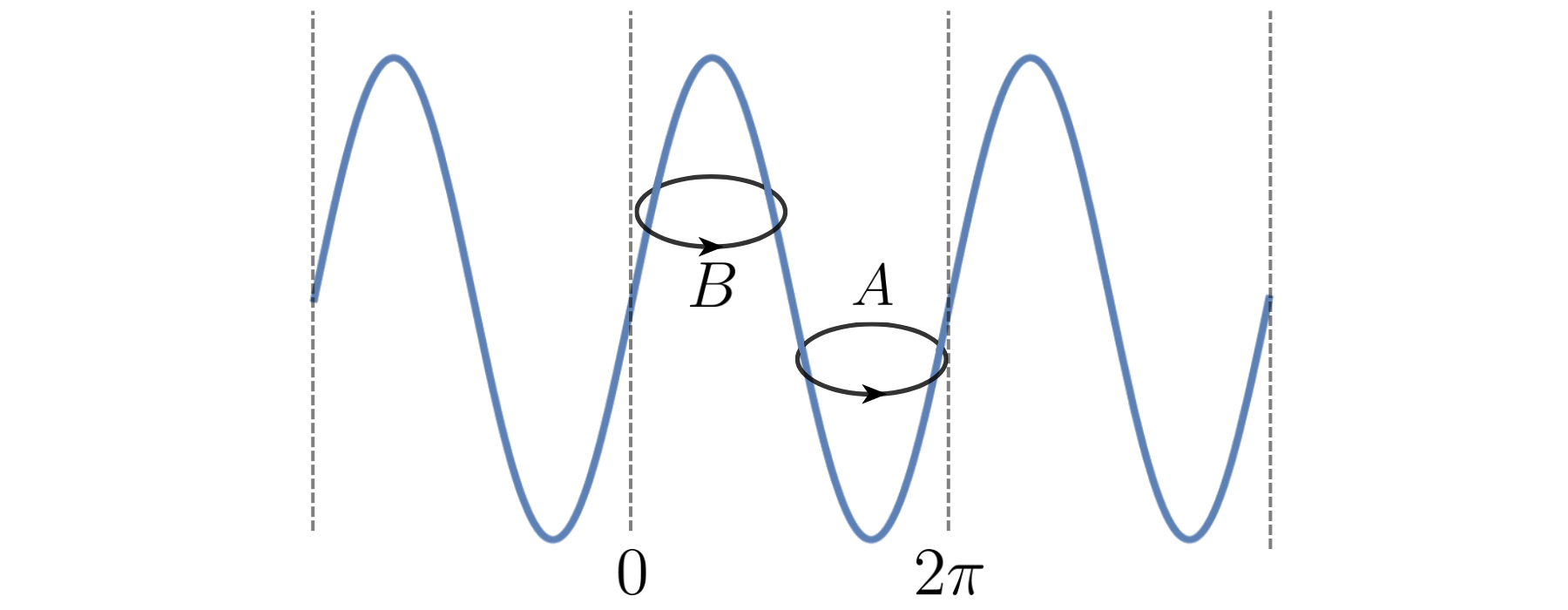}
    \caption{Periodic potential with only one distinct minimum in the fundamental domain $q\in S^1$.}
    \label{fig:cos}
\end{figure}

\begin{figure}[h]
    \centering
    \includegraphics[scale=0.15]{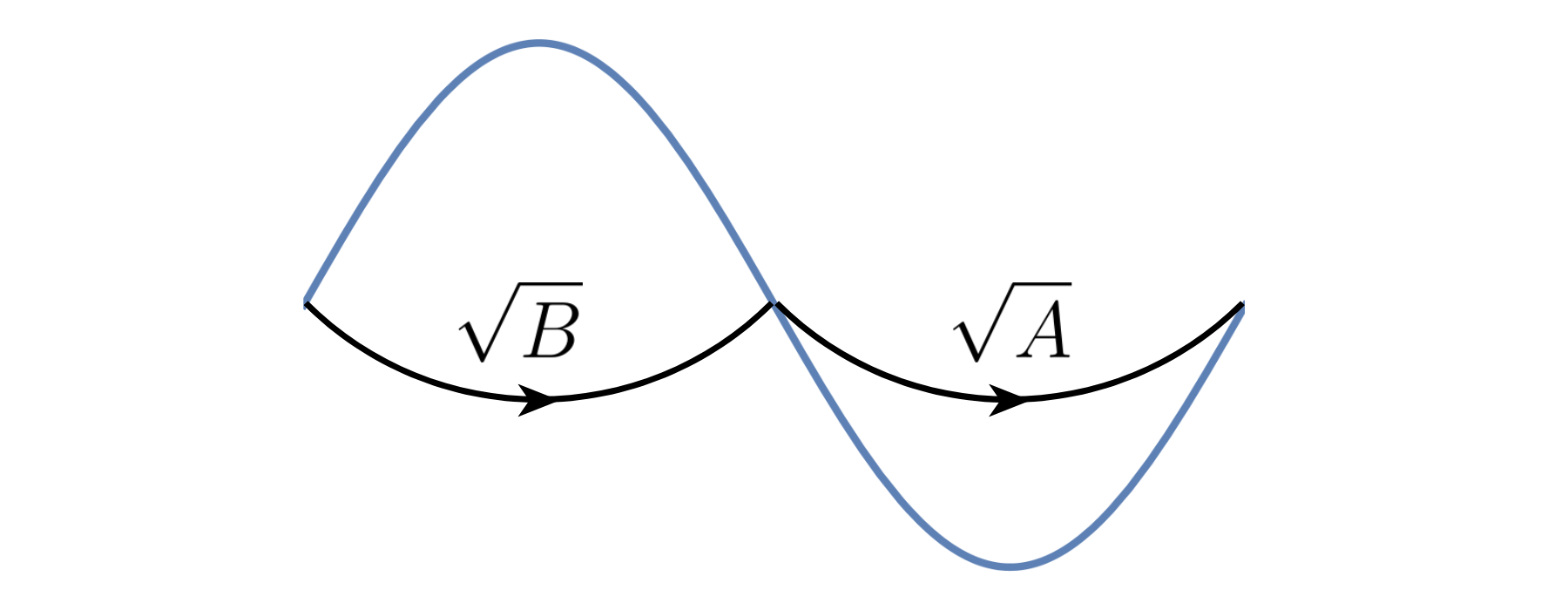}
    \caption{Topological prime periodic orbit.}
    \label{fig:cospaths}
\end{figure}

\noindent
{\bf $N=1$ minima in fundamental domain:}
For simplicity and demonstration purposes, assume that the potential $V(q)$ has only one minimum and one maximum in the fundamental domain $q\in S^1$ as shown in Fig.[\ref{fig:cos}]. This corresponds to having one classical and one dual cycle with the corresponding Voros multipliers as usual,
\begin{equation}
    A = e^{\frac{i}{\hbar}W_{\gamma}},\qquad B= e^{\frac{i}{\hbar}W_{\gamma_d}}.
\end{equation}
To do the summation over the prime periodic orbits, we need to account for the fact the points $q=q+2\pi$ are physically identified. Therefore, on top of the usual perturbative and nonperturbative minimal transmonomials
\begin{equation}
    \Phi_{\text{p}} = A,\qquad \Phi_{\text{np}}= \frac{B}{1+A}
\end{equation}
we also have the topological transmonomial that will contribute to the nonperturbative minimal transmonomial
\begin{equation}
    \Phi_{\text{top}} = -\frac{\sqrt{A}\sqrt{B}}{(1+A)}e^{i\theta} - \frac{\sqrt{B}\sqrt{A}}{(1+A)}e^{-i\theta} = -2\frac{\sqrt{AB}}{1+A}\cos\theta
    \label{topological}
\end{equation}
corresponding to another prime periodic orbit as shown in Fig.[\ref{fig:cospaths}]. This can be observed from the fact that
\begin{equation}
    \int_a^bpdq = -\int_b^apdq=\frac{1}{2}\oint_{\gamma_{ab}}pdq.
\end{equation}
where $\gamma_{ab}$ is the cycle encircling the turning points $a$ and $b$ in the complex q-plane. The minus sign in equation \eqref{topological} is from the fact that we encounter 2 turning points along the trajectory.

Hence, we can write down the spectral determinant via spectral path integral (summation)
\begin{align}
   -\log D(E) &= \sum_{n=1}^{\infty}\frac{(-1)^n}{n}\left[\Phi_p^n+\left(\Phi_{\text{np}}+\Phi_{\text{top}}\right)^n\right]\\\nonumber
    -\log D(E)&= \sum_{n=1}^{\infty}\frac{(-1)^n}{n}A^n + \sum_{n=1}^{\infty}\frac{(-1)^n}{n}\left(\frac{B}{1+A}-2\frac{\sqrt{AB}}{(1+A)}\cos\theta\right)^n\\\nonumber
    \log D(E)&= \log(1+A) +\log\left(1+\frac{B}{1+A}-2\frac{\sqrt{AB}}{(1+A)}\cos\theta\right).
\end{align}
It then implies that
\begin{align}
    \implies D(E) &= 1+A+B-2\sqrt{AB}\cos\theta = 0
\end{align}
is the exact quantization condition for the system with a periodic potential having one minimum and one maximum on the fundamental domain $q\in S^1$, \cite{sueishi_exact-wkb_2021}. 

\bibliographystyle{utphys}
\bibliography{GutzSumRef}
\end{document}